THE CONCEPTS OF TIME AND SPACE-TIME IN PHYSICS*

Máximo García Sucre

Centro de Estudios Interdisciplinarios de la Física (CEIF)
Instituto Venezolano de Investigaciones Científicas (IVIC)
Apartado 21827, Caracas 1020A, Venezuela.

ABSTRACT

We have made a revision of papers related either by affinity or by contrast to a fundamental theory of time and space-time, a previous version of which has already been published. We show that starting from the primitive concept of preparticle and of the membership relation $\varepsilon$ of set theory, and four postulates alluding simple concepts, the derivative concepts of time, space-time, reference frame, particle, field, and interaction between fields, are obtained. We have analyzed the problem of the "direction of time", and examined a paradox whose solution goes again the idea that the direction of time is determined by the growing of entropy prescribed by the Second Law of Thermodynamics. In our theory the direction of time is unique in each reference frame and it is an intrinsic property of time. Two classes of particles are considered as derivative concepts, and it is shown that one of them fulfills the Bose-Einstein statistics and the other one the statistics of Fermi-Dirac. We also consider the concepts of wave function and particle detector as derivative concepts of our theory. Making use of these concepts, we describe the process of localization of a macroscopic body (classic localization), and localization of a microscopic body (quantum localization). We also analyze similarities and differences between them.

INTRODUCTION

In every fundamental theory its concepts are divided in two classes. In the first one enters the so called primitive concepts. These are concepts that cannot be defined in terms of other concepts of the theory under consideration. That is why the primitive concepts are also called "indefinable". The second class, the derivative concepts, can be defined in terms of other concepts of the theory, either primitive or derivative concepts.

The primitive concepts of a theory can be seen as the more elementary supposition of existence from which we start in the construction of such theory. A paradigmatic example is provided by the Euclidean Geometry: the most ancient and perhaps the more beautiful example of a deductive axiomatic system. The concept of *point* is a primitive one in this axiomatic system. In this concern, Euclid made *gala* of an extraordinary economy of words when he refers to the concept of geometrical point as an entity without extension. The *geometrical point* is characterized by him denying that it has properties that the derivatives concepts of his geometry certainly have. For instance the figures described making use of the concepts of his geometry: those of line, plane, triangle, etc., each of which has extension are defined precisely using the primitive concept of point. What we have said also applies to non-Euclidean geometries, which are less intuitive than the Euclidean geometry. In what follows we will see the importance of this feature of primitive concepts in a fundamental theory.

One may ask: Why, in the construction of a fundamental theory, one must start by assuming primitive concepts, either implicitly or explicitly? The answer is precisely given by the French mathematician Henry Poincare in his article of 1891 on non-Euclidean geometries (1), which can be resumed in this way: The primitive concepts are necessary in order to avoid an infinite regression of concepts that successively are defined in terms of the following one in a regression without end. When we stop this regression the primitive concepts appears, which will depend on the place where we stop the regression. On the other hand, not any place in the regression is the most convenient to make a stop, since it is convenient to assume primitive concepts as simple and clear to the intuition as possible. It is by appropriately choosing the primitive concepts of a theory that the intuition enters to play a role in its construction, facilitating for instance the choice of *material referents* corresponding to some concepts of the theory. If we do not pay attention to this aspect of the construction of a theory, it will tend to appear blurring, or even mysterious, giving the impression of arbitrariness. The worse case arises when we suppress altogether the primitive concepts, since then we cannot choose simpler and intuitive concepts in terms of which describe other concepts of the theory, besides the logic problem that this represents.

We have, for example, the evolution of the most recent physical theories that seek the unification of Einstein General Relativity Theory and Quantum Mechanics. This search has lead to conclude that to treat this problem it is necessary to elaborate a pregeometry. That is to say, a fundamental theory from which we can derive the properties of space-time leading to the macroscopic space-time geometry of General Relativity on one hand, and on the other hand to the properties of the microscopic world predicted by Quantum Mechanics. Archibald Wheeler is one of the initiators of this idea in order to elaborate an unified description of the physical phenomena at all scales, from the intergalactic astronomical scale to the ultramicroscopic one of Planck (2-4). According to this author, it is necessary to construct a theoretical framework not presupposing the geometrical properties of space-time of General Relativity, but instead that these properties can be deduced from this more fundamental theoretical framework to which one must not ascribe geometrical properties. In the article "Is Physics Legislated by Cosmogony?" Patton and Wheeler (4) put forward five arguments to abandon the idea that the most elementary concepts of physics must be of geometrical nature. In his more recent proposal, Wheeler argues in favor of a pregeometry starting from the primitive concept of *bit*. If we take this term in his literal meaning, the argument of Bunge (5) against it is justified. According to this argument, the concept of *bit* presupposes the existence of macroscopic systems such as computers. Therefore in order to deduce the space-time properties at the microscopic level, we start from a primitive concept which depends on the existence of macroscopic objects having extension and temporality, and being of a great complexity. In other words: in order to describe the physics of the universe, to any scale from the extreme microscopic level to the extreme macroscopic one, use is made of a *pregeometry* that starts from a primitive concept that presupposes complex material objects having properties such as extension and temporality, which precisely are derivative concepts that we expect to obtain from the *pregeometry* under consideration. On the other hand, if we consider the word "bit" with quotation marks, we could be speaking about something different from a *bit* of a computer, something that could be specified either as a primitive concept with an intuitive meaning, or as a derivative concept in the framework of a given fundamental theory no presupposing the concepts of extension and temporality. In general, the ideal of a fundamental theory in which time, space-time, particle, state of a particle, interaction between particles, field associated to a system of particles, interaction between fields, are derivative concepts, and also material referents to these concepts can be specified, occurs when its primitive concepts are intuitively simple and no presupposing derivative concepts of the theory.

In the theory considered here, particles and systems of particles are represented by sets, and when we speak about particles belonging to a given system, we will not always specify whether we are referring to a physical entity or to the set representing it. Yet, it must be clear that these two entities are different: the first one is a concrete object, and the second one is a concept that represents it (5).

As an example, let us consider a theory of time that starts from the primitive concept of material object. Given the properties that we usually ascribe to material objects, such as the possibility of changing of state with time, having extension, to interact with others material objects, etc., we will be in a case of presupposition of the concept of time, which is precisely what we want to elucidate as a derivative concept. When we speak of a usual material object, we are presupposing the concept of change, since any material object of our experience, sooner or later will change from one to another state. Furthermore, the concept of change is associated with the concept of time in the following sense: in order to observe a change of state of a material object we must at least consider two different instants of time, since otherwise a material object could have two different states at the same time in a given reference frame. On the other hand, we may assume here that an instant is a finite lapse of time sufficiently short in order that no change of state occurs for the material object under consideration. Also, it is true that we tend to identify time with the way by which we measure it using periodic movements. This lead to see the progression of time as a continuous line and an instant as a point of this line. It is a kind of "spatialization" of time. We will see that despite the profound relation existing between space and time in actual physical theories, these two concepts cannot be represented in the theory developed here by mathematical entities having the same structure with respect to a given order relation to be specified later on.

Considering again our argument in connection with the relation between change and time, it is not worth to argue that according to quantum mechanics a system of particles can be found in a mixture of states in the same instant, because before performing on the system a measurement, the system will be in one state which is precisely described by the mixture considered, and not in any of the states entering in the mixture. It is only when we perform a measurement that the system may appear in *one* of these states, according to the probabilities prescribed by the coefficients weighting each term in the mixture.

Therefore, according to our argument, a theory that starts from the primitive concept of material object in the usual sense would not be a fundamental theory of time. Yet, this problem could be solved if we consider as primitive concept that of a *sui-generis* material object, with properties not presupposing space and time. This alternative will be explored in the next section.

Let us consider another example of presupposition of the concepts of time and space. In Penrose's theory (6, 7) enters as primitive concept that of "twistor", which in turn presupposes the concept of angular momentum. The classical concept of angular momentum uses the concept of mass motion with respect to an axe of rotation, and its value is calculated making use of the distance between the mass and the axe of rotation, and the velocity of the mass. Therefore, this concept is related to a reference frame with respect to which these quantities can be determined. On the other hand, in quantum mechanics there are two kinds of angular momentum: the orbital angular momentum, and the proper momentum or spin. The first kind requires of the same conceptual ingredients as classical angular momentum and the second kind needs at least the concept of reference frame. Penrose's theory also makes use of the concept of null geodesic (rays of zero intervals in the relativistic light cone). Hence, the concept of twistor presupposes the concept of reference frame, distance, and velocity, and consequently those of space and time. Yet, this theory can be seen as a fundamental theory of *matter* that has been valuated by its performance for field theory calculations that cannot be performed using other formalisms (8), although the concepts of space, time and space-time are not derivative concepts of this theory.

The case that we have just considered illustrated the fact that fundamental theories of physics are not in general fundamental theories of space-time in the sense that we have explained here. Yet the more successful physical theories, as the paradigmatic cases of General Relativity and Quantum Mechanics are not fundamental theories in the sense that we have considered here. However, these theories possess such an amazing power of prediction that there is no known fundamental theory of space, time or space-time that can compete with them in this concern. And this seems to be a characteristic feature that differentiates these two kinds of theories. On one hand, we have the fundamental theories in which one tries to start from as simple concepts as possible not presupposing the concepts that one pretends to elucidate as derivative concepts, but still these theories have little or no power of prediction. On the other hand, we have physical theories in which neither there is a clear distinction between primitive and derivative concepts, nor clarity in the physical interpretation of some of the components of its formalism, but that in general these theories have an incomparable greater power of prediction than fundamental theories. This should not be very surprising if one takes into account that the fundamental theories of physics have a preponderant orientation towards the philosophical bases of physics, while physical theories are mainly oriented to a quantitative description of the properties of matter such how these can be considered and measured. These two kinds of theories could be seen as complementary approaches of our way of knowing. The fundamental theories giving a framework to the knowledge of the physical world expressing it in a way that allows to clarify concepts and its relations to material objects, and also relating it to other fields of knowledge, and in general with the cultural system. And the physical theories non properly fundamental improving the prediction of experimental results at the cost of making less clear its logical structure and the intuitive interpretation of the material referents of the theory, and its relation to other fields of knowledge. Assuming this perspective we partially avoid the conflict between these two ways of seen our knowledge about the physical word.

An example to illustrate the remaining of some important difficulties respect this concern appeared when the elaboration of a theory unifying General Relativity and Quantum Mechanics has been attempted. In fact, several attempts have taken place to solve this important and difficult problem by developing new theories in which one presupposes the concept of space, time, and space-time (9-14), and therefore these three concepts are not elucidated as derivative concepts. These developments are important contributions, but still they do not lead to a conclusive answer to the problem of unification of general relativity and quantum mechanics. With other perspectives, string theory is one of the more successful attempts of the mentioned unification. However, in the developments of this theory have appeared interpretation difficulties of whose solution seems to require a strong change of our ideas about space-time making use of a fundamentalist approach (15). This is an indication that it may be necessary to elaborate a theory in which the geometrical properties of space-time could be obtained from simpler concepts not presupposing geometrical properties (2-4, 15). This would be a conceptual system with its respective material referents, fulfilling the requirements that we have already expressed as necessary to qualify as a fundamental theory of space-time. Clearly, this is an extremely difficult task of research in theoretical physics and foundation of physics, and what happened is that theories of space, time, and space-time approaching to the ideal of a fundamental theory of physics, with special emphasis in philosophical issues, have been proposed (16, 17).

In other words, we have physical theories extremely valuables according to their power of prediction of experimental results, but less valuables as fundamental theories of physics, and vice versa. This may be seen as a sort of complementary relation between these two kinds of theories, that has resulted in practice, and that could be tentatively enounced in the following way: More a physical theory has a greater power of prediction of experimental results, it fulfills less conditions to qualify as a fundamental theory, and vice versa. At this point, two possibilities arise: (i) There exists a reason of principle for the above mentioned complementary relationship that must fulfill our knowledge of the physical world; (ii) The mentioned complementary relationship that has resulted in practice is only one stage of the historical evolution of the fundamental theories of physics.

Let us make clear at this point that the Newtonian Mechanics is a physical theory that for macroscopic scales and velocities not approaching the velocity of light, and for masses not too large in comparison with the earth mass, has an amazing power of prediction. In addition, it is a theory that can be formalized in such a way that a clear distinction between derivative and primitive concepts can be made (16). However, its primitive concepts enfold macroscopic entities, as for instance the mass of the bodies of our direct experience, time and space. Therefore, this theory cannot be considered a fundamental theory, neither of space and time, nor it has power of prediction at microscopic scale. The same is valid for General Relativity, although in this theory neither there is restriction on the velocity of physical bodies until the limit of the velocity of light, nor on the mass values involved (excepted enormous masses as those arising at the Planck scale). Other physical theories that describe macroscopic properties of matter can be formalized in the same way as Newtonian mechanics (16, 17). Yet, these theories cannot be considered fundamental theories of physics as we are considering them here, due to the restrictive scope of its capacity of description of the physical word, and also because these theories do not elucidate among others the concepts of space, time, and space-time as derivative concepts. In this last concern, the case of Quantum Mechanics goes along the same lines despite its amazing power of prediction of the properties of matter at the microscopic level, to which it must be added that a controversy persist on its physical interpretation (16, 17).

Coming back to the possibility (i) mentioned above, at the present time there is no formal proof sustaining the mentioned complementary relationship. In any case, if the possibility (i) results to be the correct one, it might be expected that there exists an optimal balance such that a theory fulfills as good as possible the conditions to qualify as fundamental theory, having at the same time the less possible deterioration of its power of prediction. *Mutatis mutandi* that reminds the *Complementarily Principle of Bohr*, who adopts it to the extreme of considering complementary concepts, so much that once to the question posed to him by a journalist on the complementary concept of *truth,* he immediately answer: *clarity.* Despite of this, at this stage of our theoretical research we will consider valid the above mentioned possibility (ii). We pretend to overcome, at least partially, the complementary relation that in practice has resulted until now, according to which more a theory approaches an accurate description of the physical reality, especially concerning the microscopic word, more obscure tend it to be for its understanding. This is consistent with the point of view stressed here concerning the conditions that must fulfill a fundamental theory, among others the clarity of the concepts and semantic rules relating them to material objects. Although this might make difficult to get a high degree of power of prediction of experimental results, we assume here that there is no reason of principle avoiding to overcome this difficulty. In other words, let us say that it might be fruitful to think that the optimal point mentioned above is compatible with the elaboration of a theory having important traits of a fundamental theory of space-time, and at the same time having an acceptable good power of prediction.

THE CONCEPT OF TIME

The notion of time has been a subject of inquire by a handful of thinkers, many of them following a philosophical approach. Some have paid special attention to the nature of time as a physical variable, particularly in what concerns the problem of irreversibility. Others have paid attention to the psychological meaning of time, among others.

We start by examining how are related the notion of time to that of change, from a point of view related to physics. To this end let us consider the notion of state of a physical system in a given reference frame. Change occurs in a physical objects of our experience, and this is generally described as a transition of a state to another of the physical object under consideration. It is no worth to say that there are physical objects that do not change, because waiting a sufficiently long period of time surely a change of state will occur. We have extracted the notion of change from our experience. Notice, that we have said "waiting a sufficiently long period of time" for that a change of state occurs, which is an indication that the notion of time is related to that of change of state. In order to see this neatly, let us imagine that time does not exist in the sense that there is no something like two different instants of time, and that there exists only one instant corresponding to a perennial present. This would lead us to the conclusion that if we maintain the existence of change of states, a physical object could have two or more different states at the same instant in a given reference frame, and neither from a physical point of view nor from a philosophical one, this can be accepted. Consequently, if we accept the notion of change of state of a physical object, we are led to accept the notion of time as a collection of different instants, such that we could say that a physical object being found in the state $s$ at the instant $i_s$ passed to the state $t$ at the instant $i_t$. Two different states of a physical object imply two different instants of time in a given reference frame. However, two different instants of time do not necessarily correspond to two different states. This will depend on whether the lapse of time is sufficiently long or not for the occurrence of a change of state. In other words, if the changes of state are something real in physics, in the same way time is also real in physics.

Another subject is the concept of psychological time (18). According to recent researches on the nervous system, it seems that the brain can modify the temporal perception in order to adapt it to the preservation of life (18). When the brain registers a visual signal, which propagates at a higher velocity than a sound signal, if these two signals correspond to the same object and are not too much separated in time, the brain is capable to synchronize the perception of the two signals in order that they appear to be coincident. It seems that this facilitates the emergency decision making in the case that for instance the object emitting the signals be a hunger lion. Moreover, in an images succession, if one of them is particularly impacting for the observer, this image will be perceived as persisting for a longer time than the other images. This being so, it must be concluded that the brain can change the rhythm of our temporal perception (18). Yet, this does not imply that time can be reduced to an illusion created by our brain, because the brain itself pass for different states according to its physiology. And again, if time does not exist, the brain could be found as a whole in several different states at the same instant of time corresponding to a perennial present. Consequently, in order that the brain physiology be meaningful it is necessary the existence of time as something real as a succession of different instants. An idealist posture at the extreme of assuming that everything is an illusion created by the mind, will lead us to accept that even the brain is an illusion, and therefore the brain could not be the organ allowing us to have the "dream that we are a reality existing in a real world". Among other consequences, this will entail that science would be a useless game that makes it impossible (16, 17). In addition, let us argue that General Relativity is not compatible with an idealist philosophical position concerning space and time. According to this theory the mass of a body produces a modification of space-time by changing its curvature, which implicitly assumes that space and time are real physical entities, entangled to each other to give rise to space-time. We might also say that it would be very difficult to explain, in the framework of an idealist philosophy, the change of curvature of space-time produced by the presence of mass, since it would be then necessary to accept that the mass of a body modifies the way in which mind perceives space and time.

The above arguments incline us to favor that time has an objective physical reality, which does not means that we can see it as a river that flows uniformly. For instance, according to Special Relativity the observation of the same phenomenon leads to differences in the temporal lapses between the same events observed in different reference frames.

In the literature have appeared several fundamental theories of time considered as a physical variable, among them the theory of Noll and Bunge (19-21). This theory starts from the primitive concept of event and considers the case of universal time (19). It was further developed by Bunge to treat the case of local time introducing the concept of reference frame which makes it compatible with relativity (19-21). Others fundamental theories of space and time are analyzed in reference 16.

Lorente proposes a relational theory of space-time emphasizing geometric properties by postulating a cubic n-dimensional lattice where each point is related to 2n different points and only to them. In this theory one elucidates the concept of point, line, surfaces as derivative concepts by means of the relational character of the lattice. The author shows that this theory is compatible with the formulation of Hilbert of fundaments of geometry (22-24). A clear exposition of the actual fundamental theories of discrete space-time has been given by Lorente in reference 25.

In the following sections we continue to develop a fundamental theory of space-time whose preliminary stages have been already published (26-31). This is a theory of discrete space-time that is relational only in the sense that the disappearance of matter entails the disappearance of space-time. In other words, space-time is not an independent reality from matter, which makes this theory relational in a weak sense (31). This is in agreement with a classification introduced in reference 31, according to which there are space-time theories that are relational either in a weak or in a strong sense. A space-time theory is relational in a strong sense if it considers space-time only as a network of relations between material objects (31), while the theory developed here considers that space-time has a material reality.

We start from few primitive concepts that in addition should be as simple as possible. We have elucidated as derivative concepts those of event, particle, and state of a particle and of a system of particles, space, time, field and interaction between fields, reference frame, energy and momentum of particles (26-31). We consider two primitive concepts: those of *preparticle* and membership relation of set theory (26). Preparticles are considered the most elementary entities of matter. According to our discussion in the Introduction concerning the conditions that must be fulfilled by a fundamental theory, the concept of preparticle should not presuppose any derivative concepts that we pretend to elucidate in our theory. Consequently, the concepts of change of state, of spatial extension, temporal duration should not apply to preparticles, even if these concepts emerge in our theory as a consequence of the way in which preparticles conform structures that can be represented by ordered sets whose elements are subsets of preparticles. Such sets represent a type of particle in our theory, which can cross each other depending on the intersection between sets of preparticles. These intersections allow us to construct what we have called points of crossing, which may have different structures depending on the sets representing particles and the way in which they cross each other. These concepts of particle and point of crossing are examples of derivative concepts constructed starting from the primitive concept of preparticle. By now we only say about preparticles that they are a *sui-generis* form of matter, that give rise to structures that can be represented by ordered sets whose elements are subsets of preparticles.

Even if the concept of preparticle may seem a little strange, it has some similitude to the concept of event as it is considered in both special and general relativity. In these theories space-time contains all the events that occur in the universe, each of which can be differentiated from all the others. To the events are ascribed coordinates of space and time which depend on the reference frame considered in such a way that the same event may have different coordinates when it is considered in different reference frames. Thus, the coordinates of events may change but the events themselves do not change. The concepts of change and extension do not apply to events. These two traits of events, and that events can be differentiated from each other, are also fulfilled by preparticles. However, there is a difference between these two concepts since an event may be seen as "something that happens to something", while we consider that preparticles are sui-generis material entities. This difference inclines us to consider a preparticle as a primitive concept, and the concept of equivalent class of points of crossing having the same structure as a derivative concept representing the elements of space-time. Each one of these elements corresponding to a different structure, as in the same way in relativity the events differentiated from each other as elements of space-time. A comprehensive discussion of the concept of event in special and general relativity is given in reference 32. Therefore, the analogous concept in our theory of the concept of event in relativity is the point of the space-time represented by an equivalent class of points of crossing having the same structure.

It is not applicable to say that a preparticle has a big or small size, or an infinitely small size, since the concept of size does not apply to preparticles. We ascribe to preparticles neither duration nor the possibility to change of state. We ascribe to them a sui-generis material referent, extending in this way what we consider matter according to our experience about the macroscopic world. The preparticles are considered by us as a form of matter not having the usual general properties to which we are familiar, mainly through our observation of macroscopic systems. From an intuitive point of view we could say that preparticles constitute a kind of primordial soup of sui-generis material entities that are indirectly related to particles that can be experimentally detected. Yet this relation is different from the traditional atomistic relation of being components of particles. In a forthcoming section we will consider two kinds of particles: one of them such that their states are represented by sets of preparticles (bosons), the other kind corresponding instead to cuts or rips in space-time (fermions) (31). The property that preparticles have in our theory is that of yielding assortments that can be represented by sets.

In order to further clarify distinctive traits of our theory, let us make a comparison of it with two well known general conceptual frameworks (a clear description of the historical evolution of the concepts of space and time can be found in references 15, 33-35). One of these conceptual frameworks refers to Atomistic Theories of Matter, according to which every chunk of matter is composed of material atoms, taking into account different versions of what is considered to be an atom. The other general framework is the Leibniz Theory of Monads, which has a predominant philosophical character (36, 37). The contrast of our theory with these two conceptual proposals could facilitate to understand the concepts considered here. In the case of atomistic theories of matter, it is clear that our theory cannot be considered as such in the traditional sense, since in our theory we consider a subclass of particles which are simply cuts or rips in space-time (31). On the other hand, in the philosophical Leibniz theory, despite the spiritual character of monads, they are considered to be the truly atoms of nature, from which all compounds are made. Instead, in our theory preparticles enter in space-time in a way not allowing interpreting them as components of particles, in particular in the case already mentioned of particles that correspond to cuts in space-time. In the theory of Leibniz monads do not have extension. The argument of Leibniz is that if monads do have extension they could not be simple entities. This suggests that in this theory to do not having extension means to have an extension equal to zero, as if monads were points. Instead for preparticles the concepts of extension and duration are deprived of sense. In the theory of Leibniz the monads are spiritual entities, each one representing a "point of view in the universe". Instead, preparticles are material entities with sui-generis properties in contrast with the properties of matter according to our experience of the macroscopic word, and also in contrast with the properties of the microscopic word, let say to the level of components of the atomic nuclei, according to experimental high energy physics. The monads do not interact with each other. It is said that "monads do not have windows", they do not communicate among them. Due to this trait of his theory Leibniz is constrained to introduce the idea of the pre-establish harmony in order to be in agreement with the order that we observe in nature. Instead, the concept of interaction has no sense for preparticles, since this concept is a derivative concept of our theory having sense only for fields, particles, and systems of particles, even if preparticles manifest themselves indirectly in the way as fields and particles interact (26-31).

Many of philosophical inquires about time insist in two ideas: (i) Time is only an illusion; (ii) The nature of time can only be intuitively perceived (for a critical discussion of these ideas see references 16 and 33). The point of view that we develop here is the opposite of these two ideas. In agreement with the arguments given above, we consider that time has an objective reality, having properties that can be analyzed scientifically. In particular, making use of the properties of time it may be explained some experimental observations about physical properties. Furthermore, if starting from the primitive concept of preparticle, whose material referent is a sui-generis form of matter, and from the derivative concepts of time and space-time, one elaborates a theory that gives a coherent explanation of experimental observations that have not be explained until now, or have been explained in a controversial way, this would be an indication of the existence of preparticles. Clearly, this is precisely one of the objectives more difficult to attain.

Let us starts by the collection of all preparticles in the universe, and we represent this collection by means of the *base set* whose elements are all the preparticles (26):

$$B= \{ \alpha_1, \alpha_2,... \},\qquad(1)$$

In reference 29 we consider some arguments in favor that the number of preparticles in the universe is finite, although amazingly large. In the next section we explain how time can be elucidated as a derivative concept of our theory, and similarly the concept of reference frame. To every reference frame corresponds a way for ascribing temporal coordinates to the points of space-time. In this context, a schematic way to represent time can be described as follows (26). Starting from the *base set B* of preparticles we construct the power set *P(B)* whose elements are all the subsets of *B*. Any pair of these subsets may be either disjoints, or partially intersecting to each other or still one of them included in the other. Let us consider now a collection of subsets of the set *B* included in each other in such a way that all of them remain ordered by the inclusion relationship. In this way we construct a mathematical representation that orders completely these subsets of preparticles by using only the membership relation of set theory to decide which preparticles belongs to each of the considered subsets. If we assume in a preliminary simplified interpretation that each one of these subsets of preparticles represents an instant of time, then we obtain in a constructive way starting only from sets of preparticles a representation of time as a succession of instants. Note that each one of these instants does not flow: each one does not change and belong to a collection whose elements are represented by subsets of preparticles, such that these subsets are ordered by the inclusion relationship (26). The concept of time associated to a reference frame may be seen *grosso modo* in the following way: a reference frame may be constructed by selecting a set of space-time points and particles connecting these points. Among these particles there are some of them whose elements are represented by subsets of preparticles, and when they are ordered by the inclusion relation, this gives rise to what we have called *evolutive particles*. These particles may connect space-time points allowing ascribing temporal coordinates to these points. Analogously, considering particles that in turn connect the mentioned evolving particles we may also ascribe spatial coordinates to the considered space-time points. Different ways of selecting particles that connect space-time points give rise to different ways to ascribe coordinates to these points, and thus to different reference frames (27, 28) (see for instance figure 2).

In this simplified way of representing time we have not presupposed the concept of change. The assumptions that we have made are: (i) the existence of prep articles as a form of sui-generis matter; (ii) the representation of flocks of prep articles by sets; (iii) the representation of a succession of instants of time, by subsets of prep articles ordered by the inclusion relation.

FIELD PRODUCED BY A SYSTEM OF PARTICLES

The sets whose elements are subsets of preparticles can be classified in two kinds according to their structure with respect to the inclusion relationship. As we have already mentioned some of these sets can be completely ordered by the inclusion relationship, and others do not. In the first case these sets represents particles that we have designate as *evolving particles*, and in the second case the sets represent particles that we have designated as *non evolving particles* (26).

Before going on, let us introduce the concept of *alpha-state* of a particle. For the sake of clarity, let us consider the case of an evolving particle whose graphic representation is illustrated in figure 1a. This representation appears as a continuous curved line on which are successively marked points A, B, C, D, E, and F. Given that we have considered that the total number of preparticles is finite, although enormously large, let us point out that in figure 1a we have assumed that each molecule of paper over which pass the curve AF corresponds to a preparticle. In this way we avoid that each preparticle corresponds to a point of a continuous curve, which would not be consistent with our assumption that the total number of preparticles is finite. Notice that the elements AB, AC, AD, AE and AF of the particle represented by the curve AF, are successively included in one another according to: $AB \subset AC \subset AD \subset AE \subset AF$. We call alpha-state of the particle illustrated in figure 1a the curve segments AB, BC, CD, DE and EF. Notice that the order of the alpha-states represented by the segments AB, BC, CD, DE and EF is not based in the graphic representation of the considered particle. In our theory, the order and structure of the particles and its alpha-states can be specified making use only of the elements of the theory without using any figurative procedure. Yet, in our illustration of figure 1a we have made use of such figurative procedures with an illustrative purpose.

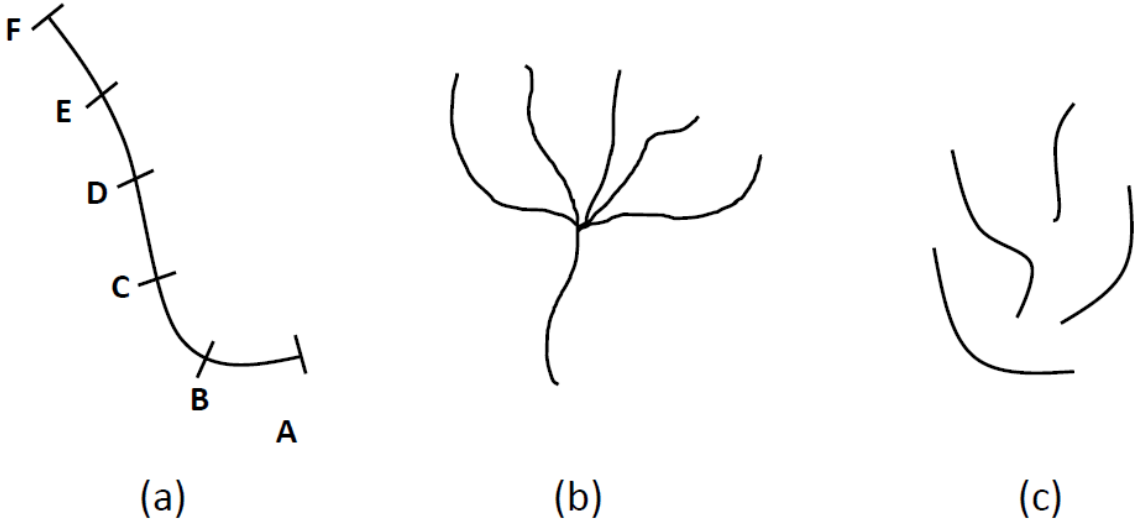

Fig 1: (a) Evolving particle whose elements are represented by the subsets AB, AC, AD, AE, and AF. The alpha-states of the evolving particle are represented by AB, BC, CD, DE and EF. (b) Non evolving particle having five branches. (c) Non evolving particle having four branches

In a more formal description we start from the *base set B* of eq. (1) and define sets representing particles and alpha states of particles in the following way (26-31):

$$p_i = \{ a^i(x) \mid x \in X \text{ and } a^i(x) \in P(B) \}, \qquad (2)$$

Where the set $p_i$ represents a particle, $P(B)$ is the power set of B minus the empty set, and the alpha-states of $p_i$ are given by:

$$s^i(x) = a^i(x) - \cup a^i(x'), \qquad (3)$$

Where $a^i(x)$ is an element of $p_i$, and the $a^i(x')$'s are all the elements of $p_i$ that do not include $a^i(x)$. In the case that $p_i$ represents an evolving particle, one may consider intuitively that $a^i(x)$ corresponds to a stage of evolution of the particle $p_i$, and each $a^i(x')$ is a precedent stage to $a^i(x)$ of the evolution of the particle $p_i$.

Figures 1b and 1c illustrate two cases of non evolving particles. These figures were elaborated following the same conventions as in figure 1a. In figure 1b appears a ramified curve, and in figure 1c four curves that do not intersect each other. It is clear that figures 1b and 1c corresponds to two different cases of non evolving particles, since in each case the elements of the particles represented cannot be completely ordered by the inclusion relation ⊂. Examining figure 1b, we see that there are several different cases that corresponds to the same figure. One case corresponds to five evolving particles having a subset of elements which is the same for the five particles, yet each of them also has different elements from the elements of the other four particles. Another case would be of two evolving particles both having a subset of elements a, and three particles having a subset of elements that coincides only partially with *a* (a particular case would be that of the three particles having only one element that coincides with only one element of *a* in the point of ramification). In the first case we have five evolving particles when we consider each of them separately, but when considered together they correspond to a non evolving particle, since its elements cannot be completely ordered by the inclusion relation ⊂. We will call evolving branch of this non evolving particle each one of the five evolving particles considered in the first case. The second case also corresponds to five evolving particles that all together can be considered as one non evolving particle, one of them being an evolving branch whose first element do not coincide with the first element of the other four particles. Following the same procedure we may consider a sizable number of different non evolving particles all of them corresponding to a ramified graph identical to the one illustrated in figure 1b. Consequently, the description power of the order relations based on the inclusion relation ⊂ among subsets of preparticles, allows discriminating different cases of ordering that one cannot discriminate using graphic representations as those of figure 1b. In the case of figure 1c, each curve represents an evolving branch of the non evolving particle having four branches, given that each of these four branches can be completely ordered by the inclusion relation ⊂. In both cases illustrated in figures 1b and 1c we speak about *evolving branches* of non evolving particles,

Let us consider a set S whose elements are sets that represents the particles $p_i$ with i = 1, n, where n is a very large number. The set S will represent a system of a very large number of particles. Two particles of S will intersect one another if at least one alpha-state of one of these particles yields a non empty intersection with at least one alpha state of the other particle. Making use of the same conventions as for the figure 1, we could obtain a figure in which appears a very large number of curves representing the system S, some curves intersecting each other, and some other curves do not intersecting, which would give to the illustration of the system S the structure of a network of intercrossing curves. This network could have regions more or less dense in curves, and other regions where there are no curves, and thus corresponding to cuts or rips of the network. Let us consider now, the point where curves cross each other. These points may be such that few particles crossover them, while other points may be such that in them many curves crossover. We will call *center of a point of crossing the alpha-state where curves crossover, and these curves filaments of the point of crossing.* More precisely, we represent a point of crossing as an ordered pair $(s^i(x); \prod_x^i(S))$, where $s^i(x)$ represents an alpha-state of a particle of the system S, and $\prod_x^i(S)$ is the set of all sets that represents particles of S having alpha-states yielding non empty intersections with $s^i(x)$. This last set of preparticles represents the center of the point of crossing, and $\prod_x^i(S)$ is the set of its filaments. A particular case of point of crossing occurs when only one filament starts or ends at the center $s^i(x)$. The simplest point of crossing occurs for only one particle having only one alpha-state. Let us denote as $\Sigma\Sigma(S)$ the set of all points of crossing of the system S.

*We say that two points of crossing belonging to $\Sigma\Sigma(S)$ have the same structure when we can put in correspondence the centers of these points of crossing, and the filaments of one of them can be put in a one-to-one correspondence with the filaments of the other point of crossing, in such a way that alpha-states of two filaments in correspondence have a similar ordering.*

Notice that we have not mentioned others crossing that may present the filaments of a given point of crossing. The only crossing of curves that accounts for a given point of crossing occurs at its center. The others crossing that may present its filaments will give rise to other points of crossing. We denote by the symbol ~ the relation between two points crossing having the same structure.

We represent the field C associated to a physical system S by the cocient set $\Sigma\Sigma(S) / \sim$. The elements of the set $\Sigma\Sigma(S) / \sim$ represent the points of the field C. This way of representing the field associated to a system S of particles have the following properties (27-31):

(i) To a system S of particles corresponds a field represented by $\Sigma\Sigma(S) / \sim$.

(ii) Each element of $\Sigma\Sigma(S) / \sim$ represents a point of the field associated to the system S of particles. For each equivalence class x of points of crossing we say indistinctly that a point of crossing belongs to x or enters in x when this point of crossing has the structure corresponding to the equivalence class x.

(iii) By definition of equivalence class, the points of a field represented by the elements of $\Sigma\Sigma(S) / \sim$ are such that each point have a different structure, and therefore each point of a field singularizes from all the others points by its structure.

(iv) The points of a field C represented by $\Sigma\Sigma(S) / \sim$ may be connected by particles of the system S, in such a way that given a particle *p* of S having at least one alpha-state yielding a non empty intersection with the centers of two points of crossing entering in two points of C, then we will consider that these two points are connected by the particle p. Similarly, we say that two particles p and p´ are connected by a particle p´´ if alpha-states of p´´ yield non empty intersections with alpha-states of p and p´.

(v) Two points x and y of a field C represented by $\Sigma\Sigma(S) / \sim$ are next to each other if there exists and evolving particle p (or an evolving branch of p) of S connecting x and y in such a way that at least an alpha-state of p yields non empty intersections with centers of point of crossing entering in the points x and y, or else the alpha-state of p that yields a non empty intersection with the center of a point of crossing entering in the point x is the immediate antecessor or the immediate successor of the alpha-state of p that yields a non empty intersection with the center of a point of crossing entering in the point y.

(vi) Given a set $\Sigma\Sigma(S) / \sim$ representing the field associated to S, we define uniformity of field making use of the concept of field intensity in a given point x of a field C, which is the number of preparticles that belong to the union of the centers of points of crossing entering in x. A field C is uniform if the field intensity is the same in all its points.

Two fields C1 and C2 associated respectively to the systems S1 and S2 interact with each other if the following inequality is fulfilled: $\Sigma\Sigma(S1 \cup S2) / \sim \neq (\Sigma\Sigma(S1) / \sim) \cup (\Sigma\Sigma(S2) / \sim)$ (27). It can be shown that two fields C1 and C2 do not interact with each other if the systems S1 and S2 do not share particles and also no particle of S1 crossover a particle of S2. It can be also shown that C1 and C2 interact with each other when particles of S1 crossover particles of S2 in such a way that at least one of those crossings give rise to a point of crossing with a different structure from those of the points of crossing entering in the points of C1 and C2. It may also occur that even in the case that particles of S1 crossover particles of S2, still C1 do not interact with C2. This happens when points of crossing arising from crossings of particles of S1 with particles of S2 give rise to point of crossings having the same structure as points of crossing arising uniquely from particles of S1 or from particles of S2.

The above discussion on the interaction between fields fits well the intuitive idea of superposition of fields, either when there is interaction between them or not. In order to conclude our discussion on the concept of field let us introduce the following postulates (28):

Postulate 1: *"Any region of the physical word is formed by points of fields or by points of superposition of fields".*

Postulate 2: *"Two regions of the physical word are different from each other only because the points of fields or of superposition of fields that enter in one of these regions have different structures from the structures of the points that enter in the other region.*

SPACE-TIME AND ITS EXTENSION

Let us consider now a collection of systems S1, S2, ..., $S_i$ and the corresponding fields $C_1, C_2, ..., C_i$ associated to these systems and represented respectively by $\Sigma\Sigma(S1) / \sim$, $\Sigma\Sigma(S2) / \sim$, ....., $\Sigma\Sigma(S_i) / \sim$. *To every collection of fields $C_1, C_2, ..., C_i$ associated respectively to the systems S1, S2, ..., $S_i$, we associate a space-time represented by ST (S)= $\Sigma\Sigma(S) / \sim$, where S= S1 U S2 U....U $S_i$.* Therefore, in our theory, given several fields, either interacting or not between them, the space-time associated to these fields can be intuitively seen as the global or resulting field. The elements or points of a space-time are represented by equivalence classes of points of crossing, and these elements play a similar role to the one played by the concept of event in Special and General Relativity, as we have already considered. With this definition of space-time it is possible to specify reference frames in a space-time. This can be done in such a way that given a reference frame we may ascribe without ambiguity space and time coordinates to each point of space-time, and also that the coordinates ascribed to these points may change when considering different reference frames (27, 28).

Before describing the way in which a reference frame may be specified in order to ascribe coordinates to the points in a given space-time, let us first analyze an important property related to its extension. The simplest idea that suggests itself concerning the extension of a discrete space-time, is that it is related to the number of its points, and in a certain way is what we will assume. Yet, the extension of a space-time ST(S) or of a field C in our theory may be very far from a proportionality relation to the number of particles entering in the system S and to the number of alpha-states of these particles. In fact, the extension of ST(S) depends of what we will call the complete character of the system S. We understand by a complete system with respect to a subset B´ of the base set B, a system

S to which belongs all the particles that can be represented by subsets of the power set P(B´), and only that particles. In this case we will also say that the space-time ST(S) is complete. This property is related to two theorems that we have shown in reference (30), and that can be enounced as follows: Given a space-time ST(S), where S is complete with respect to a given B´ included in B, and if we only consider the particles of S in order to form the points of crossing that give rise to the points of ST(S), then this space-time will have only one point, and therefore its extension reduces to zero.

Here we consider this property in a less formal way than in reference (30). We start with the following definitions:

Definition 1: A space-time is of order n if all the alpha-states of the particles of S are sets of n preparticles, and we denote it ST(S, n).

Definition 2: A space-time ST(S, n) is complete of order n with respect to a subset B´of the base set B, when to S belongs all the particles, and only these particles, that are represented by subsets of the power set of B´, such that alpha-states of these particles are all sets on n preparticles.

Then we have the following properties:

1) An immediate property is that ST(S, n) has only one point when n is the total number of preparticles of B´. In this unique point enters only one point of crossing of only one particle having only one element that coincides with its unique alpha-state of n preparticles (see equations (1)-(3)).

2) Let us consider another possible extreme case when n=1. In the case that B´ is a set of a very large number of preparticles, ST(S, n) may be a space-time in which enters a very large number of points of crossing, but only when ST(S, n) is far to be complete it could have a large number of points represented by equivalence classes of these points of crossing. In order to see this, let us consider that ST(S, n) is complete and that x is a point of it. Given that n=1 the centers of all points of crossing entering in x are represented by sets of only one preparticle. Let us now assume that there exists a point x´ of ST(S; n=1) different from de *x*. Then in x´ enters points of crossing having a different structure than the points of crossing entering in x. Moreover, the centers of the points of crossing entering in x´ should be alpha-states to which belong preparticles different from those belonging to the centers of the point of crossing entering in x, since two different points of crossing cannot have the same center (they would form together only one point of crossing having this center). Consider now two points of crossing a = (s$^i$ (x); $\prod^i_x$(S)) and b = (s$^j$ (y); $\prod^j_y$(S)) that enter in the points x and x´, respectively. Let us consider a particle p of the point of crossing *a*. Then there exists an alpha-state *s* of *p* that has a non empty intersection with the center s$^i$ (x). According to our assumption that the space-time ST(S, n) is complete, all the possible subsets of the power set of B´ corresponding to alpha-states of only one preparticle represent particles of S. Thus, in the point of crossing *b* will enter a particle *p´* of S with the same alpha-states of *p* except the alpha-state *s* that now is an alpha-state *s´* which yields a non empty intersection with the center s$^j$ (y) of the point of crossing *b*. Furthermore, we can chose *p´* in such a way that there exists a one-to-one correspondence preserving the order of the alpha-states of *p* with respect to those of *p´*, and that puts in correspondence the alpha-states *s* and *s´*. In this way we can construct particles that enter in the point of crossing *b* that are similar to the particles entering in the point of crossing *a*, in such a way that the centers of *a* and *b* are in correspondence. Since we can follow the same procedure now starting from the particles of *b* in order to construct the particles of *a*, we conclude that the points of crossing *a* and *b* have the same structure, and therefore the point *x* and *x`* of ST(S, n) have the same structure. Now, according to the Postulate 2 the only feature that allows to distinguish two point is their structures, which leads us to conclude that every complete space-time ST(S, n=1) has only one point. As a consequence we may expect that in the measure that ST(S; n=1) approaches to be complete it will have fewer points, independently of the fact that the set B´ may have a very large number of preparticles. Following an analogous procedure it can be seen that ST(S, n>1) also reduces itself to a point when it is complete. In order to prove this property one follows the same steps as for the case n=1, but taking into account that the intersection between alpha-states of the particles may now be only partial in contrast with the case n=1, in which two alpha-states intersect completely or not at all.

   3)Let us consider two subsets B´ and B´´ of the base set B of all preparticles, such that to these subsets belong the same number of preparticles, and let S´ and S´´ system of particles that can be represented by sets that can be constructed starting from B´ and B´´. If the sets ST (S´, n) and ST (S´´, n) represent complete space-times, then by the property (2) both of them reduce to have only one point. Let us denote respectively *x´* and *x´´* these points. The points of crossing entering in *x´* will have the same structure as the points of crossing entering in *x´´*. As a result, the points *x´* and *x´´* are not different from each other since according to postulate 2 two points are different only by its structure, and consequently ST(S) with S=S´U S´´ will have only one point. Notice that in this case ST(S) is not a complete space-time, since we may construct sets of subsets of preparticles starting from the power set P( B´U B´´) that do not represent particles of S. Then ST(S) reduces itself to having only one point despite the fact of not being a complete space-time.

   4) The more general case arises when we do not introduce any restriction concerning the number of preparticles belonging to each alpha-state of the particles of S. Also in this case a complete space-time reduces itself to have only one point. This property can be proven following similar steps as we have considered above in the more simpler case of a complete space-time ST(S, n), but now it must be taken into account that to alpha-states may belong different number of preparticles, and that two alpha-states may intersect partially. In spite of these complications it can be proven that if ST(S) is complete then it reduces itself to having only one point. A proof to this property in the more general case can be found in reference (30). Similarly the property (3) can be generalized to the case in which there is no restriction to the number of preparticles that each alpha -state may have.

   5) In the case of a partitions of the base set B of all preparticles, in such a way that every element of the partition has the same number of preparticles, it can also be proven that the complete space-times that can be constructed starting from the elements of the partition of B, all of them reduces itself to having only one point, and all these points have the same structure. On the other hand, in the case of subsets of B having non empty intersections among them, whose union is equal to B, and each of them have the same number of preparticles, then we get a similar result as for the case of the partition of B that we have already considered: the space-time $ST_e(S, n)$ such that in S enters particles represented by all the sets that can be constructed starting from the considered subsets of B, then $ST_e(S, n)$ reduces to have only one point. Thus, despite that in the considered case $ST_e(S, n)$ is not complete, it also reduces to having only one point, and $ST_e(S, n)$ is not complete, since we can construct sets starting from the power set of B that do not represent any particle of S. In fact, the particles of S are only those that can be represented by the sets that can be constructed starting from the power sets of the considered subsets of B. In consequence, we see that the strategy of partitioning B, either in subsets having empty intersections or not, in a kind of mosaic of subsets of B, each having the same number of preparticles, do not allow to obtain space-times having more than one point. However, we will see below that when the subsets of B do not have the same number of preparticles, either having empty or not empty intersections, the space-time $ST_{ee}(S)$ that can be constructed according to the procedure already described may have more than one point. In fact, it may have a very large number of points.

6) Let us consider the subsets B´ and B´´ of the set B of all the preparticles, and let n´ and n´´ be the numbers of preparticles of B´ and B´´, such that n´ > n´´. It is clear that in the power set of B´ we will have subsets that cannot be constructed starting from the power set of B´´, since some representations of particles corresponding to B´ will have more alpha-states that any representation of particles constructed starting from the power set of B´´. Therefore, some points of crossing that can be constructed starting from B´ cannot have the same structure than the points of crossing that can be constructed starting from B´´, and thus the structure of the unique point of ST (S´) is different from the structure of the unique point of ST (S´´), where S´ and S´´ are complete systems whose particles are represented by subsets of the power sets of B´ and B´´, respectively. On the other hand, given that the number N of all the preparticles is very large (29), then also may be large the number of points having different structures that may be constructed starting from subsets of the base set B. In particular, this occurs when the union of these subsets is equal to the base set B of all the preparticles, and to each of these subsets belong a number of preparticles which differs strongly from the number of preparticles entering in others of these subsets, all these number being very small in comparison with N.

OBSERVATION SYSTEM IN A SPACE-TIME

Let us now see that there exists a way of representing a field, or more generally a space-time, that differs from those that we have already considered in 1)-6). It is a way more nearly approaching the procedure entering in play in a physical observation. Despite the fact that generally we do not enounce it explicitly, in classical physics we consider that the observer has no influence in the physical system under observation, and accordingly we may consider that these two entities are completely separated from each other. In the classical description of physical systems the influence of the observer on the observed system is erased by the way in which classical physics describes physical systems. Classical physics does not include in its conceptual ingredients the concept of observer. In spite of this, classical physics gives an excellent physical description of macroscopic systems, even so we know that the macroscopic systems are also quantum systems, and that the observer clearly interacts physically with the observer system (for instance, electromagnetic and gravitation interaction). Yet, the effect on the observed system, both due to its quantum nature as due to the physical interaction with the observer, is almost always negligible for macroscopic systems. Consequently, in classical physics we can ignore the observer and still give a very precise physical description of macroscopic systems.

Here we keep aside of any interpretation of the observer as an entity with psychological attributes. We shall not enter in the controversy that still subsists in connection with the concept of observer in quantum physics (for instance, the problem of the conscience of the observer in the detection of a quantum event). We will consider that the influence of the observer on an observed system of particles is produced only by physical interactions. Consequently, instead of the term observer S2 of the system S, we make use of the term *observation system S2 of the system S*.

In the quantum description of microscopic systems, the system S2 has an influence on the observed system S, and this influence is such that it cannot be ignored. From the beginning of quantum theory this has been considered in many ways giving rice to different interpretations to the influence of the observer on the observed system (17, 35). Experiments inspired in the *gedankenexperiment* ("thought experiment") proponed by Einstein, Poldowsky and Rosen (EPR) (38), have been undertaken in the last 30 years (32, 39, 40). These experiments have rather give ground to the orthodox interpretation of quantum mechanics concerning the influence of the observer on the observed system. This result carries with it a certain irony, given that the "thought experiment" EPR was conceived to demonstrate that the orthodox version of quantum mechanics is an incomplete theory, which was not confirmed by the mention experiments. Yet, it is also proper to say that the concept of observer in the orthodox version of quantum mechanics enters in the conceptual structure of the theory as an ad hoc aggregate in order to facilitate the interpretation of the experimental results (16, 17). Attempts have been made to modify the formulation of quantum mechanics in order to introduce the concept of observer in the core of the theory in equal status to the concept of observed system (35, 41). Here also we will try to include the concept of *Observation System S2* in the conceptual structure of our theory at the same level than the concept of *Observed System S,* in such a way that both kinds of system are considered as physical systems of particles, where each particle is represented by sets of subsets of preparticles.

Brief, in classical physics the effect of S2 on S is ignored, although it is taken into account in quantum physics. One way to handle this feature is to consider that S2 is included in the system S. There is no problem with this assumption concerning classical physics, since in this case the influence of S2 on S can be ignored. However, in quantum mechanics the influence of S2 on S must be taken into account, giving rise to various interpretations that consider different ways in which S2 relates to S (15, 35).

*POSTULATE 3: Let us consider a system S of particles represented by the subsets of the power set of B´, P(B´), where B´ is a subset of the base set B of all the preparticles. Consider also two subsystems S1 and S2 of the system S = S1 U S2 and S1=S ▬ S2. Then the system S2 is an observation system of S and the extension of the space-time ST(S) with respect to S2 is given by the number of points of ST(S1).*

In order to make clear the concept of *system of observation* let us consider the following example. Let the space-time ST(S) be complete, which implies that each element of S is represented by a subset of P(B´), where B´⊂ B, and that there is a one-to-one mapping between S and P(B´). In this case, the space-time ST(S) has only one point, as we have already argued, and the extension of ST(S) with respect to S2 will be given by the number of points of ST (S1). Now, depending on the particles belonging to S2 the extension of ST(S) with respect to the system of observation S2 may be a number different of one, and even it may be a very large number, depending on the number of preparticles of B´ and on the particles belonging to S2. Let us assume that to B´ belong a very large number of preparticles. This implies that the number of points of crossing belonging to the unique point of ST(S) will also be very large. All these points of crossing have the same structure and thus they all belong to the unique point of ST(S). If S2 is a system of only one particle $p_1$, and to the point $x$ of ST(S) belongs a very large number of points of crossing, the majority of the points of crossing will not be affected by omitting the particle $p_1$ from S, and consequently the extension of ST(S) with respect to the observation system S2={ $p_1$ } will corresponds to a relatively small number of points: one of them still having a very large number of points of crossing, and a few others to which belong a small number of points of crossing. Now consider the following opposite case: to S2 belong almost all the particles of S. This implies that to S1=S—S2 belongs a small number of particles and thus a few number of points of crossing will enter in each point of ST (S1). Then, this space-time will have a small number of points, since this number cannot be larger than the number of points of crossing that can be formed with the particles of S1. Let us now describe an intermediate case between the two cases that we have just considered. We assume that to the system S2 belongs a large number of particles, still much smaller than the number of particles belonging to S. Then the extension of ST(S) with respect to the observation system S2 could be large, since the structure of a large number of points of crossing belonging to the point x of ST(S) could be modified by omitting from S a large number of particles (all those belonging to S2). This will occur except in cases where the particles of S2 are selected in a very special way. For instance, when the particles of S2 modify the points of crossing that belong to $x$ always in the same way, and therefore the structure of the modified points of crossing will be the same for all of them, and again all of them will belong to only one point of ST(S1). On the other hand, if to S2 belongs a large number of particles randomly chosen it would be also a large number of different ways of choosing the particles that belong to S2 and that would give rise to a large extension for ST(S) with respect to the observation system S2, in contrast with what happens when we chose the particles of S2 in such a way to modify in the most similar possible way the points of crossing belonging to the point $x$ of ST(S). Since we do not know which are these particles when we consider a given macroscopic system S2 as an observation system of ST(S), S2 having a large number of particles, although much smaller than the number of particles of S, then the extension of ST(S) with respect to S2 could be very large, and it will be so with respect to almost any macroscopic S2 randomly chosen.

    Brief: 1) If S2 is a system having a very small number of particles, then independently of the number of particles of S, the extension of ST(S) with respect to the observation system tends to be small ; 2) One has the same result when S2 ⊂ S is near to have the same particles as S ; 3) The largest extension of ST(S) with respect to S2 is generally obtained when S2 is a system with a very large number of particles, but still with a number of particles much smaller than the number of particles of S.

Postulate 4: "*The Universe is formed by all the particles of the system S that can be represented by subsets of the power set P (B), where B is the set of all preparticles, and its extension depends on the observation system S2 considered, and is given by the number of point of ST (S1), where S=S1US2 and S1=S—S2.*"

SPACE-TIME AND REFERENCE FRAME

Once we have introduced the concept of observation system S2 of a given system S, we now describe different ways in which space and time coordinates can be ascribed to the points of ST(S1), where S1=S—S2. First, consider one of the simplest ways to define reference frames (30). Given a subset $\tau$ of evolving particles belonging to the system S1, one selects those points of the space-time ST(S1)= $\Sigma\Sigma$(S1) / ~ that are connected by particles belonging to $\tau$. Consider also a subset $\sigma$ of evolving particles of S1 such that each of them connects particles belonging to $\tau$. Then each evolving particle *p* belonging to $\tau$ will induce a partial or complete ordering on the subset of points of space-time ST(S1) such that alpha-states of p have non empty intersections with centers of points of crossing belonging to the points of ST(S1) (when this occurs we say that particle p cross-over these points of ST(S1)). This induces the ordering of these points according to the ordering of the alpha-states of *p*. In this way we can ascribe time coordinates to these points of ST (S1). This can be done without ambiguity when each particle *p* of $\tau$ cross-over once or none with a point of ST (S1). In this case, the time coordinate of each of these points could be the ordinal of the corresponding alpha-state of p. In this case the points of ST(S1) with which *p* cross-over are completely ordered, and we could ascribe to one of these points the time coordinate 0 (origin), and to the other points ordered according to the order of the alpha-state of the particle *p* time coordinates ...-3,-2,-1,0, 1,2,3,...etc. By following this procedure we can ascribe as many time coordinates as alpha-states have *p*. On the other hand, if each alpha-state of p has a non empty intersection with points of crossing belonging to several points of ST (S1), then the induced ordering by the alpha-states of *p* would be a partial ordering. Consequently, in that case each of these time coordinates would be ascribed not to one point but to a subset of points of ST (S1).

When the particles belonging to $\tau$ do not intersect each other in ST (S1), we then may consider them as oriented parallel segments that are in turn ordered by the particles belonging to the set $\sigma$. In order that there is no ambiguity in the way in which coordinates are ascribed, it is necessary that the particles belonging to $\sigma$ be such that all of them induce the same ordering of the particles belonging to $\tau$. The simplest way to do it occurs when to $\sigma$ belongs only one evolving particle p´ that intersects each particle belonging to $\tau$ in only one point (figure 2). Then we can ascribe the time coordinate 0 to all these points. Similarly, we may choose a particle belonging to $\tau$ and ascribe to it the space coordinate 0. Then to ascribe one space coordinate to each particle belonging to $\tau$ in agreement to the ordering induced in $\tau$ by the unique particle p´ belonging to $\sigma$. In this way we establish a procedure to ascribe coordinates of space and time to a collection of points of ST(S1), and by so doing we have defined a reference frame inside this space-time. The case that we have just considered corresponds to a reference frame with one spatial dimension and its corresponding temporal dimension. Let us denote R (ST (S1), $\tau$, $\sigma$) to this reference frame.

A reference frame may cover only partially the points of a given space-time, which depends on the particles of S1 belonging to the sets τ and σ. In principle these sets may be chosen in such a way that space and time coordinates could be ascribed to all the points of ST (S1).

We have described above a simple case of a reference frame defined inside a given space-time (30), which is illustrated in figure 2. There are others ways to do it, as is considered in references 27 and 29. A more natural way to define a reference frame is making use of systems of macroscopic bodies. This would be more nearly related to the way in which reference frames are established in practice. This idea will be explored below when we analyze the concept of macroscopic body in the framework of the present theory.

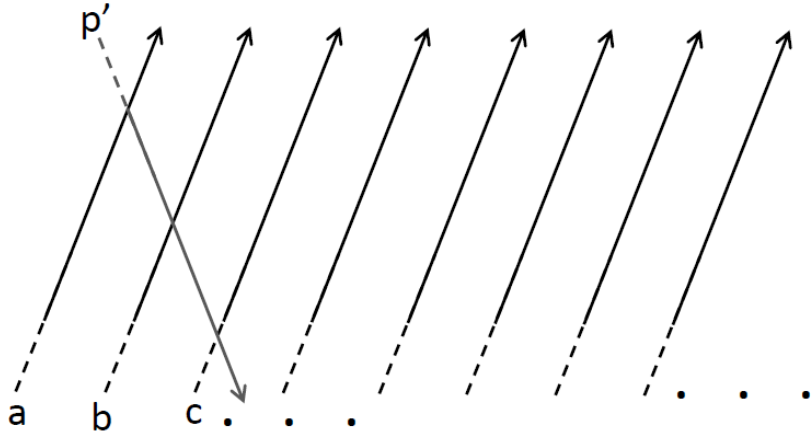

Fig. 2. The lines *a, b, c,*.... represent particles belonging to the set τ.
*p´* represents the unique particle that belongs to the set σ.

Consider again our definition of space-time as a global field corresponding to a collection of fields. The properties of fields also apply to space-time. In particular, to the concept of space-time applies the postulates 1 and 2 already stated. The elements (points) of a given space-time are formed only by particles, and when we consider all the particles of the universe we will speak about space-time as one unique entity. Given the systems S2 and S, the space-time ST (S1) = ΣΣ(S1) / ~, with S1= S—S2, may have a fragmentary aspect, with few points, presenting cuts (or rips) and eventually non homogeneities in field intensity (especially when S1 is a system of few particles). It may also occurs that depending on S and S2 the space-time ST(S1) may have an aspect intermediate between the above described fragmentary case and the more smoother case already discussed (28). We make the distinction between connection and homogeneity of a given field, and also of a space-time. A field is connected if it does not present cuts or rips in the network of its points. A field is homogeneous when its intensity is the same in all its points.

The track of a particle *p* in a space-time ST(S) with respect to the observation system S2 is represented by the set of points of ST(S, S2) that are intersected by the particle *p*, which we denote h (ST( S, S2), p). We define trajectory of a particle $p_j$ in a reference frame

R(ST(S, S2), *τ, σ*) as the set of space and time coordinates, ordered according to the increasing value of the temporal coordinates of points such that the alpha-state of $p_j$ have non empty intersections with centers of points of crossing belonging to these points. This trajectory will be denoted T(R(ST(S, S2), *τ, σ*), $p_j$ ). The track of a particle in a space-time ST(S, S2) is a partial or completely ordered set of points that generally appear sparsely in ST(S, S2). On the other hand, the trajectories of particles are partially or completely ordered sets of space-time coordinates of some points of given reference frames. Sometimes we will speak about the points of certain trajectories, in the understanding that we are referring to the points that precisely have the space and time coordinates that belong to this trajectory.

The sets of points corresponding to tracks of two particles in ST(S, S2) may intersect each other. Let τ be a set of evolving particles of S1=S-S2 such that their tracks do not intersect each other, or equivalently that the sets representing them are disjoints. Also let σ be a set of evolving particles of S1 such that any particle *p´* belonging to σ either intersect in only one point the track in ST(S, S2) of particles belonging to τ, or do not intersect them at all. If, on the other hand, each alpha-state of particles belonging to τ has a non empty intersection with centers of points of crossing belonging to one point of ST(S, S2), then the track in ST(S, S2) of each particle belonging to τ will be a completely ordered set. Furthermore, the sets of tracks  h( ST(S, S2), $p_1$), h( ST(S, S2), $p_2$),..., h( ST(S1), $p_k$), of particles belonging to τ, ordered by intersection with particles of σ, may be a set of tracks such that one of them connect with either none, one or several tracks of particles belonging to τ. We say that the connection of two tracks of particles of τ is between neighbors when the particles of σ that connect these tracks are such that one or two consecutive of their alpha-states connect them. Given a particle $p_k$ belonging to τ, we call order of consecutive connection of $p_k$ in the reference frame R(ST(S, S2), *τ, σ*),  to the number of particles of τ whose tracks are  connected as neighbors with h( ST(S, S2), $p_k$), and we denote it as  Oστ(h( ST(S, S2), $p_k$)).

An ideal reference frame, which we denote RI (ST(S, S2), *τ, σ*), must fulfills the following conditions:

(j)The tracks h (ST(S, S2), $p_k$) of particles $p_k$ of *τ* in a space-time ST(S, S2), are represented by completely ordered sets.

(jj) The order of consecutive connection Oστ (h (ST(S, S2), $p_k$)) is the same for all tracks h(ST(S, S2), $p_k$), except for the borders of the reference frame.

When the order of consecutive connection of each particle $p_k$  is 2, the corresponding reference frame is spatially one-dimensional. Due to our assumption according to which the number of preparticles is finite, if Oστ (h (ST(S, S2), $p_k$)) is 2 it should be at least two particles $p_k$´s such that their order of consecutive connection is 1. Therefore, the mentioned reference frame would be one-dimensional spatially with borders of dimension 0. The borders of the temporal axis always have dimension zero. The origin of the temporal coordinates is conventional and may be different from borders of the temporal axis. More generally, we will consider reference frames with orders of consecutive connection respectively equal to 2n and 2n-2.

The points of a reference frame R (ST(S, S2), *τ, σ*), are the points of ST(S, S2) to which spatial and temporal coordinates have been ascribed by the way in which they are connected by the particles belonging to $\tau$ and $\sigma$. The trajectory T(R(ST(S, S2), $\tau$, $\sigma$), $p_j$) of a particle $p_j$ of S in a given reference frame R(ST(S, S2), $\tau$, $\sigma$), is the ordered set of space-time coordinates of the points of this reference frame, such that a part or all the alpha-states of $p_j$ have non empty intersections with the points of crossing that belong to these points of R(ST(S, S2), $\tau$, $\sigma$). The partial or total ordering of the elements of the trajectory T(R (ST(S, S2), $\tau$, $\sigma$), $p_j$) is induced according to the growing order of the temporal coordinate. This procedure to ascribe coordinates is compatible with the usual way in which reference frames are used in physics, since the ordering of the elements of the trajectory of the particle $p_j$ in the reference frame R is defined according to the increasing values of the temporal coordinates. In this way, independently of which could be the ordering of the alpha-states of $p_j$, the ordering of the trajectory of $p_j$ in R is such that it corresponds to increasing values of the temporal coordinates. Notice that in a given reference frame R(ST(S, S2), $\tau$, $\sigma$), the trajectory of a particle of S1 may appear in R as a set of space-time coordinates such that some of them may correspond to the same time coordinate, and consequently we have in this case that the elements of the trajectory are only partially ordered. Notice also, that the considered particle p might be a non evolving particle, but still it may have a completely ordered trajectory in R which will depend on the temporal coordinates in R. The other way around, we also may have that the considered particle p may be an evolving particle, and still this particle may have a trajectory in R only partially ordered. These different cases arise because, to the exception of the particles belonging to $\tau$ or to $\sigma$, which define the reference frame considered, the elements of the trajectory of one particle are ordered in agreement with the temporal coordinates that are determined by the considered reference frame.

NON INERTIAL REFERENCE FRAMES AND GRAVITATION

According to our discussion above a reference frame will be inertial when all its points have the same intensity of field, or equivalently when the number of preparticles belonging to the union of the centers of the points of crossing entering in one point, is the same for all the points of the reference frame (27). Let us consider a randomly chosen particle $p$ of S1=S-S2 in the reference frame R (ST (S, S2), $\tau$, $\sigma$). The points of the track of the particle $p$ will tend to appear in R without accumulations of points in small regions of R. Since the particle have being chosen at random the probability that one of its alpha-state share preparticles with a given point of the reference frame R is proportional to the number of preparticles that enter in the point of R. In reference 28 we have made use of this argument to show that if we represent photons as sets defined in equation 2, the mean velocity of a sufficiently large number of photons is the same in all reference frames. According to this, if we consider light rays corresponding to large number of photons, the velocity of light will be a constant c in any inertial reference frame. Starting from this point, if in addition we require to be linear the transformations of space-time coordinates in passing from one inertial reference frame to another, one obtains the Lorentz transformations. In reference 29 it is also discussed how frequency and wave length can be defined for photons represented in the described way.

On the other hand, when the reference frame R (ST(S, S2), $\tau$, $\sigma$) is not inertial, and the number of preparticles entering in the points located, let us say, in a narrow stripe around the trajectory of a particle belonging to $\tau$, then the trajectory of a particle p randomly chosen will tend to appears as a set of coordinates corresponding to an accumulation of points in the mentioned stripe. Moreover, when the non inertial character of the reference frame R corresponds to a distribution of field intensities of the points of R described by a gauss distribution centered at one of the temporal axes of R, then the trajectory of *p* will tend to appear curved toward this temporal axis. This suggests a connection between gravitation and the non inertial quality of a reference frame in our theory (29). Moreover, if we now define a non inertial frame R´ that covers only a stripe of points of R around the trajectory of *p* and this stripe is very narrow in comparison to the width of the Gauss distribution above mentioned, then the trajectory of *p* in R´ will tend to appear as if this referential is inertial. Again this is in agreement with the way in which one considers the relation between non inertial reference frames and gravitation making use of local reference frames.

If we do not consider any reference frame within a space-time ST(S, S2), the track of a particle *p* may correspond to points of this space-time, but then it could not be possible to ascribe space-time coordinates to these points. In this case different regions of ST(S, S2) can be identified only by the way in which its points are connected; points whose structure differ from one another according to the definition of point of a given field, or of a given space-time.

MACROSCOPIC SYSTEMS IN A SPACE-TIME

Let us consider again a space-time ST(S) whose extension is defined with respect to a given observation system S2. The extension of ST(S) with respect to S2 is given by the number of points of ST(S, S1) ≡ ST(S, S2), where S1US2=S and S1=S—S2. We say that two points *x* and *y* of ST(S, S2) are *immediately connected* if there exists at least a particle *p* of S1 that enters in some crossing point *pcx* of *x*, and that also enters in a point of crossing *pcy* of y, in such a way that the alpha-state of p that has a non empty intersection with the center of *pcx* has also a non empty intersection with the center of *pcy*, or it is the immediate antecessor or successor of the alpha-state of *p* that yields a non empty intersection with the center of *pcy*. Consider in addition that ST(S, S2) has a very large number of points, and that there exists a subset STM(S, S2) of ST(S, S2) fulfilling that in each of its points enters a large number of points of crossing, in contrast with the rest of the points of ST(S, S2).

Now let z be a point of STM(S, S2). This point will be immediately connected with a large number of points of STM(S, S2). In order to see it, consider a particle *p* of S1 that enters in a point of crossing *pcz1* of *z,* and let α1, α2 , α3,... be alpha-states of *p*. Assume in addition that α2 is the alpha-state of *p* that has a non empty intersection with the center of the point of crossing *pcz1*. Given that *p* belongs to S1 it would be a point of crossing *pcz´1* that belong to a point *z´* of ST(S, S2) in which *p* enters, and let say that the alpha-state α3 of *p* has a non empty intersection with the center of *pcz´1*. In our theory the number of alpha-states of any particle is finite since the total number of preparticles is also finite. Then the point of crossing *pcz´1* will necessarily have a different structure from the structure of the point of crossing *pcz1,* even when the only difference existing between these two points of crossing comes from the fact that two different alpha-states of p are those which intersect their centers. Consider this to be the case. The point's *z* and *z´* of ST(S, S2) have different structures and consequently are two different points of this space-time. In addition, all the point of crossing belonging to a given point of a space-time have the same structure, and therefore the change introduced by the difference in the alpha-state of *p* that cross-over the centers of *pcz1* and pcz´1 should correspond to the same modification of the others points of crossing belonging to *z* in order to give rise to the point *z´*. In other words, if the particle *p* enters in a point of crossing belonging to z, it should be in each point of crossing belonging to z at least a particle similar to *p*, and the modification of the alpha-state of p that cross-over the center of *pcz1* may be considered also for all those particles similar to *p*, this giving rise to points of crossing entering in z´. Furthermore, the point z´ constructed in this way will have the same number of points of crossing than those entering in *z,* and thus z´ will qualify to enter in the subset STM(S, S2) of points of ST(S, S2) having a large number of points of crossing. In addition, the point's *z* and *z´* are immediately connected to each other at least by the particle *p* and all its similar particles entering in the points of crossing of z and z´. Since we can follow the same procedure starting from any other particle entering in a crossing point belonging to *z,* it will be more than one point of STM(S, S2) immediately connected to the point *z*. Three cases may arise: (i) the points of crossing belonging to the points of STM(S, S2) are such that only few particles of S1=S-S2 cross each other in each of them; ii) in each one of the points of crossing enter many particles of S1; iii) intermediate cases between the cases (i) and (ii). In the first case, each point of STM(S, S2) will be connected with less points of STM(S, S2) than in the second case. The third case corresponds to a situation in which some points of STM(S, S2) are immediately connected with many points of STM(S, S2) and others points do not.

Consider now the reference frame R (ST(S, S2, $\tau$, $\sigma$) defined in the space-time ST(S, S2) in such a way that to every point of this space-time corresponds space and time coordinates. In agreement to these coordinates we can establish for each point *x* of R (ST(S, S2), $\tau$, $\sigma$), the neighboring points of this reference frame to the point *x*, both spatially and temporally. Let SC1 be the system of particles of S1 that fulfills with the condition of entering in points of crossing belonging to the points of STM(S, S2). Then, we say that the system SC1 of particles is a macroscopic body in the space-time ST(S, S2), if for any reference frame R that ascribes coordinates to all the points of this space-time, the particles of SC1 crossover the points of R in such a way that a spatially wide region R´ of R much larger than what corresponds to the spatial separation between two neighbor points of the reference frame, and temporal duration covering all the temporal axis of the reference frame, for any point x´ of R´ there is at least a particle of SC1 entering in a point of crossing that belongs to x´.

In the case of a macroscopic body SC1 corresponding to the alternative (ii) described above, we have a space-time STM(S, S2) for which many particles of SC1 enter in each point of crossing belonging to a point of STM(S, S2). Also, we may expect that many points of crossing enter in each point of STM(S, S2). We have argued that for each of these points it would be at least one different point of STM(S, S2) immediately connected with it. This allows us to define what we may call a *path between two points x and x´* of STM(S, S2): A path between *x* and *x´*, which we denote C(x, x´), is a set of points of STM(S, S2) such that the point x has a point immediately connected with it that belongs to C(x, x´), let this point be x´´ that in turn is immediately connected with x´´´ which also belongs to C(x, x´), and so on and so forth until getting a point which is immediately connected with x´. So, the points of C(x, x´) constitute a chain of points such that except for x and x´ each point of C(x, x´) has an immediate predecessor and an immediate successor in C(x, x´), the point x having only immediate successor, and the point x´ having only immediate predecessor. We call x and x´ border points of the path C(x, x´). Two points' x´´ and x´´´ of a path may be immediately connected in two different ways. To see it, let α1 and α2 the alpha-states of an evolving particle p by mean of which this particle connects immediately x´´ with x´´´. Then, there are two possibilities: α1 proceed to α2 as alpha-states of the particle *p*, or vice versa. We represent the first case with an arrow going from x´´ to x´´´, and the second case with an arrow going from x´´´ to x´´ (respectively, x´´ → x´´´ and  x´´← x´´´). Thus, a path C(x, x´) between two points x and x´ of STM(S, S2) may be symbolized by mean of a set of oriented arrows:

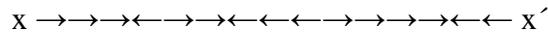

Fig.3: Path C(x, x´) of 16 points and 15 arrows in the space-time STM(S, S2) between x and x´. Each arrow represents an evolving particle.

We have illustrated in Figure 3 a path C(x, x´) in the space-time STM(S, S2) between the points x and x´, having 16 points and 15 arrows. Of these arrows,  9 are from left to right, and 6 from right to left. We define now *minimal path* Cm(x, x´) between the points x and x´ of the space-time STM(S, S2) as the path between x and x´ having a number of points equal or less than the number of points of any other path C(x, x´) between x and x´ in the space-time STM(S, S2). The spatial separation between two points x and x´ of STM(S, S2) is given by the number of pair of arrows in opposite directions occurring in the minimal path Cm(x, x´), and the temporal separation by the difference between the number of arrows corresponding to opposite directions. Let us assume that the path illustrated in figure 3 is in fact a minimal path Cm(x, x´). In this case, the spatial separation between the points x and x´ of STM(S, S2) will be of 4 units, and the temporal separation will be also of 3 units. Notice that spatial separation is always measured in positive units or zero. The sign of spatial separations arises once we arbitrarily choice as positive the spatial separation between x and x´, which in turn fixes that the spatial separation between x´ and x will be negative. On the other hand, the sign of a temporal separation is given by the sign of the difference between the number of arrows having opposite directions and will be determined once we choice one of the two possible directions as positive.  Therefore, concerning the signs of spatial and temporal separations between two points x and x´, we have that for the temporal separation the sign is given by the orientation of the arrows appearing between x and x´, and for the spatial separation the sign arises once an additional convention is introduced.

The concept of path between two points x and x´ of the subspace-time STM(S, S2) corresponding to a macroscopic body SC1, allows to ascribe coordinates to the points of STM(S, S2). As we have mentioned when we have defined the concept of reference frame in a space-time, the concepts of macroscopic body and of path between two points may be used to specify a reference frame in a given subspace-time. In order to see this, consider all the paths that can be specified in a given STM(S, S2), in such a way that all the arrows (see figure 3) be similarly oriented, for instance from left to right. In this case, the initial point and the final point of these paths are separated by arrows, all of them oriented in the same way, and then with no pair of arrows opposing to each other. Therefore, for all these paths the spatial interval will be zero. If among these paths we select now those that do not cross each other, we will be in the same situation that we had when we have defined reference frames using the sets that we have denoted $\tau$ and $\sigma$. But now, the elements of our set $\tau$ will be the paths that we have just considered and that give rise to temporal intervals, and the elements of the set $\sigma$ are paths that crossover the paths belonging to $\tau$, in the same way that we have already considered when we have defined reference frames using sets $\tau$ and $\sigma$ whose elements are evolving particles. The procedure to specify reference frames starting from the concept of macroscopic body is more nearly related to the usual procedure to select a concrete physical reference frames than other procedures that we have described.

Finally, let us consider a case complementary to the above described. Consider a system of particles SC1´ in a space-time ST(S, S2) such that for all reference frame R that ascribes coordinates to all the points of this space-time, the particles of SC1´ only cross points of R that appear disperse in a region R´ of R, or either disperse in a region covering R completely. In other words, the points of R that cross the particles of SC1´ do not cover R in a dense way. We will call SC1´ *microscopic system of particles or microscopic body*. According to reference (29), when the disperse distribution of these points in R is such that these points have a uniform spatial separation, we say that the microscopic system SC1´ has a well defined wavelength in R. Similarly, if the temporal separation between these points is uniform we say that SC1´ has a frequency well defined in R.

DETECTOR SYSTEMS OF PARTICLES

Let us assume that we have a space-time ST(S) having an extension defined with respect to the system of observation S2, which we denote by ST(S, S2), the reference frame R (ST(S, S2), $\tau$, $\sigma$) defined in this space-time, and the particle *p* having a trajectory in the reference frame R represented by the set T(R (ST(S, S2), $\tau$, $\sigma$), p). Consider also a system of particles represented by $S_d$ which in turn is a subsystem of S. According to our definition of extension we have S1US2=S and S1=S—S2. Assume that the trajectory of *p* in the reference frame R (ST(S, S2), $\tau$, $\sigma$) has the aspect of a disperse distribution of points. In agreement with our definition of microscopic body we consider as such the system of particles to which belongs only the particle *p*. Consider now the space-time ST´(S, S2 U $S_d$) and the reference frame R´ (ST(S, S2 U $S_d$ ), $\tau$, $\sigma$), which correspond to S1´= S1— $S_d$ and S2´=S2 U $S_d$, in such a way that we still have S=S1´U S2´ and S1´=S—S2´ with $S_d \subset S$. This corresponds to have added to the system of observation S2 the subsystem $S_d$ of S1.

Let us now see that it may occur that the trajectory of the particle p en the reference frame R´ may have a narrow region in spatial distribution during a time interval $\Delta t$, at difference with the aspect of the trajectory of the particle p in the reference frame R. If $\Delta t$ is larger than a threshold value $\Delta t_0$ of time interval, we say that the system of particles represented by $S_d$ is a detector of the particle p with respect to the threshold $\Delta t_0$ in the reference frame R´. In principle, the threshold value $\Delta t_0$ should be chosen in a way that the detection considered could be characterized in an experiment. The way in which may occur the mentioned narrowing in the distribution of points in passing from the reference frame R to the reference frame R´ can be understood as follows. In passing from the system of observation S2 to the system of observation S2´= S2 U $S_d$, the extension of ST(S) with respect to S2 U $S_d$ will tend to be in some cases larger than the extension of ST(S) with respect to S2, and in other cases will tend to be lesser. An extreme case will arise when S1= $S_d$, since then the extension of ST´(S, S2 U $S_{d)}$ reduces to zero: in this case the extension of ST´ (S1´, S2 U $S_{d)}$ would be the number of points of ST´ (S1´= S1— S1, S2U $S_d$) which is an empty set. Therefore, this case of detection of a macroscopic system of particles corresponds to a collapse of space-time, of the reference frame, and of the proper microscopic system of particles. In a different case to the above extreme one, let us say that to $S_d$ belongs much less particles than to S2, and in turn to S2 belongs much less particles than to S. Then the space-time ST´ and the reference frame R´ will have extension, and it may occur that the particles of $S_d$ cross some points of crossing belonging to points of R modifying its structure. This produces a redistribution of these points of crossing giving rise to the points of R´. Thus, the trajectory of *p* in R´, that is determined by the points of crossing having centers with which the alpha-states of *p* have a non empty intersection, may change of aspect when contrasted with the trajectory of *p* in the reference frame R, and thus it is possible that the above mentioned narrowing occur. Furthermore, the number of particles of $S_d$ may be so small in comparison with the number of particles of S2, that in passing from ST to ST´ and from R to R´, the number of points of these entities remains constant, and only the mentioned redistribution of points of crossing occurs. Yet, when the system $S_d$ is not so little with respect to S1, the number of points in passing from ST to ST´ will tend to increase. As in principle we do not have information about what are the preparticles that enter in the particle *p*, neither in the points of R, nor in the points of R´, given a system represented by $S_d$, we cannot be sure about the detection of *p* by $S_d$, although in principle we can speak of probability that this event occurs. This is a characteristic trait of quantum mechanics, which we will consider below in more detail when we introduce the concept of wave function in the framework of the present theory.

Consider now the detection of a microscopic body. Again, in the extreme case in which S1= $S_d$, the space-time ST´(S, S2 U $S_d$), the reference frame R´ (ST(S, S2 U $S_d$), $\tau$, $\sigma$), and the macroscopic body STM(S — S2U $S_d$), all three are to empty sets. On the other hand, if in $S_d$ enter much less particles than in S2, and also in S2 enter much less particles than in S, then the space-time ST´ and the reference frame R´ will have sizable extensions. The particles of $S_d$ will cross the center of points of crossing entering in the points of R, with the consequent redistribution of these points of crossing to give rise to the point of R´. But now, at difference to the case of a microscopic body it will not be a change in density of points of the macroscopic body trajectory in passing from the reference frame R to the reference frame R´. This may be expected on the ground that the distribution of the points of a macroscopic body trajectory is already dense in the reference frame R, and will continue to be so in the reference frame R´. The effect of $S_d$ on ST(S, S2) tends to make increase the number of points to give rise to ST´(S, S2 U $S_d$). This increase in the number of points will tend to be negligible since to $S_d$ belongs much less particles than to S2.

Other extreme case of reduction to zero of the extension of a space-time arises when S2 and $S_d$ are empty sets. In this case ST´(S, S2U $S_d$)= ST (S), and therefore ST´ is a complete space-time, i.e. only one point belongs to it and then its extension reduces to zero. This case corresponds to one kind of collapse considered in reference (30), leading to a space-time with only one point to which belong all the points o crossing that can be constructed starting from the base set B of all the preparticles, since all these points of crossing have the same structure.

WAVE FUNCTION OF A PARTICLE

Here we consider another definition of particle detector that seems to us more directly related with concepts linked to quantum mechanics. One way to do it is considering first the concept of wave function. Let us start from the space-time ST(S) having an extension with respect to the system of observation S2, which we denote ST(S, S2), and the reference frame R (ST(S, S2), $\tau$, $\sigma$) defined in this space-time. For the sake of simplicity let us consider first the case in which R has only one spatial dimension. Consider a particle *p* belonging to S, and the set S2´= S2 U {p}. The trajectory of p in the reference frame R is formed by all the space-time coordinates of the points of R with which the particle p crossover, i.e., the points of R in which enter points of crossing whose centers have non empty intersections with one or several alpha-states of p. Consider now the space-time ST´(S, S2´), which differs from ST only in that the particle p does not enter in any point of crossing belonging to a point of ST´(S, S2´). In the same way, we define R´ (ST´(S, S2´), $\tau$, $\sigma$) for which we assume that the sets $\tau$ and $\sigma$ do not change in passing from R to R´. Also we assume that the structure of the points of crossing that enter in the points of R is such that by effect of omitting the particle p these points of crossing redistribute to yield the points of R´, in such a way that this does not affect the number of points of R´. In other words, the number of particles of S ― S2´ is so large that the effect of p only produces a redistribution of points of crossing among the points of R to give rise to the points of R´. The fact that by effect of the subtraction of the particle p of the system S could produce a redistribution of points of crossing among the points of R to give rise to R´ without change of the number of points of ST´(S, S2´) with respect to the number of points of ST(S, S2), may be seen in the following way: the modification of the points of ST(S, S2) to give rise to ST´(S, S2´) by omitting the particle *p* of the system S will produce the redistribution of points of crossing already mentioned, and in order to form a new point it would be necessary that by omitting the particle p one obtains points of crossing whose structure is different to the structure of any point of crossing entering in a point of ST(S, S2). Yet, if an enormous number of particles belong to S, the number of points of ST(S, S2) could also be enormous, and as we know each of these points corresponds to a different structure of points of crossing. Consequently, what will occur more frequently is that there already exists the structures that are obtained by omitting the particle p from the points of crossing that enter in the points of ST(S, S2) to give rise to ST´(S, S2´). Thus, according to our assumption that in S2 and S2´ enter a number of particle much smaller than the number of particles of S it may be expected that by omitting the particle p only a redistribution of the points of crossing among the points of ST(S, S2) to give rise to ST´(S, S2´) occurs, the number of points remaining essentially constant.

Consider now a point x of R having the value t as time coordinate. Let us denote X (t)

the set of points of R having this value t for their time coordinate. Also denote N+(x, t) the number of points of crossing that belong to the point x in which enter the particle p, and that are such that when omitting from them the particle p they get the same structure as some or all the points of X(t) —{ x }. Also, denote N-(x, t) the number of points of crossing that belong to the points of X(t) that are such that the particle *p* enters in these points of crossing, and that by omitting the particle p they get the same structure as the point x. In more figurative terms, N+(x, t) is the number of points of crossing that "go into" the point x in the instant t due to the particle p, and that come from points of X (t). Similarly, N-(x, t) is the number of points of crossing that "go out" from the point x in the instant t due to the particle p to enter in some or all the points of X (t).

In addition consider the points of the reference frame R (ST(S, S2), $\tau$, $\sigma$) that enter in the time axis defined by the particle $\tau i$ belonging to the set $\tau$ and that crossover the point x of R. Let us denote T(R, $\tau i$) this set of points. Now, also denote T+(x, t) the number of points of crossing belonging to x in which enter the particle p, and that by omitting the particle p from these points of crossing they get the same structure as some or all the points of T(R, $\tau i$) —{ x }. Similarly, T-(x, t) is the number of points of crossing entering in the points of T(R, $\tau i$) such that the particle *p* enters in these points of crossing, and that by omitting the particle p they acquire the same structure as the point x. In a similar way as we have done for the points of X(t), we interpret in a figurative way that T+(x, t) is the number of points of crossing that "go into" x in the instant t due to the particle *p* and that come from the points of the time axis T(R, $\tau i$). Similarly, T-(x, t) is the number of points of crossing that "go out" from the point x in the instant t due to the particle p for going to enter in some or all the points of T(R, $\tau i$).

Now, we are prepared to define what we understand by wave function of a particle p in a reference frame R (ST(S, S2), $\tau$, $\sigma$) specified in a space-time ST(S, S2). We use italic *x* and *t* to denote space and time coordinates of the points x´s of the reference frame R (ST(S, S2), $\tau$, $\sigma$).

Definition 3. Given a space-time ST(S, S2) and a particle p represented by an element of S-S2, the wave function of the particle p in the reference frame R(ST(S, S2), $\tau$, $\sigma$), is given by the complex function:

$$\Psi (x, t) = \Psi r (x, t) + i\ \Psi i (x, t), \tag{4}$$

In which the real component $\Psi r (x, t)$ depend on *x* and *t* and is given by the difference N+(x, t) — N-(x, t), and the value of the imaginary component $\Psi i (x, t)$ is T+(x, t) — T-(x, t).

In order to facilitate the discussion we have considered the case of reference frames having only one spatial dimension. However, the above definition can be generalized to cases of multidimensional reference frames. In the spatially tridimensional case we will denote the space and time coordinates by italics *x, y, z, t* .

The expression (4) for a wave function of a particle has some properties similar to the properties of functions in quantum mechanics. First, let us see how the expression (4) transforms itself by temporal inversion. According to our definition of reference frame in a space-time the construction of the temporal axis of a reference frame starts from a set $\tau$ of evolving particles as we have already described in page 21. According to this definition we cannot reverse time, since the direction of time in a reference frame is determined by the inclusion relation of the elements belonging to sets that represents evolving particles belonging to the set $\tau$, and the ordering of a set induced by the inclusion relation cannot be inverted in such a way that its elements be again ordered by the inclusion relation. What can be done is to change the sign of the time coordinates in the considered reference frame. And this is what strictly corresponds to the usual operation of changing the sign of that appears in the arguments of a wave function. According to our definition 3 of wave function, the change of sign of the time variable can be interpreted in the following way: In expression (4) for the real component of the wave function

$$\Psi r\ (x,\ t) = N_+(x,\ t) - N_-(x,\ t), \qquad (5)$$

to change the sign of the variable t do not affect the right hand side of the above expression, since $N_+(x, t)$ and $N_-(x, t)$ only have to do with the number of crossing points that enter and go out the point x in a given instant t, and the change of sign of t will not affect these numbers. However, the sign of the imaginary component of equation (4)

$$\Psi i\ (x,\ t) = T_+(x,\ t) - T_-(x,\ t) \qquad (6)$$

will change, since in passing from t to -t is equivalent to read in reverse the increment in the temporal coordinate of the reference frame R. Therefore, the number of crossing points going into x and that come from the points of the temporal axis passing over x, that we have denoted $T_+(x, t)$, this number will be now the number of points of crossing that go out from x to go into the points of the temporal axis of R passing over the point x, i.e. that now $T_+(x, t)$ and $T_-(x, t)$ interchange their roles in equation 6. In other words, the change of sign of t is equivalent to read what enter in x as what is going out. We thus see that $\Psi i\ (x, t)$ will change of sign with the change of sign of t.

The way in which we have described the change of sign of t in the framework of our theory, is analogous to the way in which this change of sign is interpreted in classical physics as an inversion of the particle velocities whose trajectories are described in a given reference frame R. In our theory, "the direction of time" is fixed by an intrinsic property of the sets of subsets of preparticles, these sets representing temporal axes of a given reference frame. It is not determined by a property of many particles systems as the Second Law of Thermodynamics. Below, we discuss an argument according to which to assume that the direction of time is given by the increase of entropy with time of isolated systems leads to a contradiction.

According to some recent cosmological models, the universe started concentrated in a very small region having very low entropy, and that this entropy has increased with the expansion of the universe. This way of describing the history of the universe agrees well with the fact that in the processes that we observe of isolated or approximately isolated systems, almost always occur an increase of entropy (33). In principle, this way of describing the evolution of the universe does not deny the possibility of reversing the direction of time, but this possibility has no practical relevance. However, the problem is not only of practical character, since the idea that it is the increase of entropy what fixes the direction of time leads to a difficulty of principle. This difficulty consists in that if a decrease of entropy corresponds to a reverse of the direction of time, then time increases in this inverse direction and we will see the entropy to increase again instead of decreasing, since once the direction of time is reversed what was "before" converts itself in "after". Therefore, we could never observe a decrease of entropy, which contradicts the experimental observation that in processes of isolated and also non isolated system eventually happen fluctuations corresponding to a decrease of entropy. The solution of this difficulty is to accept that it is neither the decrease nor the increase of entropy what determines the direction of time.

We consider now a definition of particle detector making use of the concept of wave function given in definition 3 (equations 4, 5 and 6). Let us start by considering the space-time ST´(S, S2 U $S_d$ ) and the reference frame R´ (ST(S, S2 U $S_d$ ), $\tau$, $\sigma$), such that ST´ and R´ correspond to S1´= S1— $S_d$ and S2´=S2 U $S_d$, fulfilling S=S1´U S2´, S1´=S—S2´ and $S_d \subset$ S. This corresponds to subtract the system $S_d$ from S1 and to add it to S2. Now, consider a particle $p$ of S´ whose wave function in the reference frame R(ST(S , S2), $\tau$, $\sigma$) is $\Psi (x, t) = \Psi r (x, t) + i \Psi i (x, t)$, such that for a given $t$ the points x´s of R(ST(S , S2), $\tau$, $\sigma$) for which $\Psi r (x, t) \neq 0$ and/or $\Psi i (x, t) \neq 0$, appear disperse in the reference frame R. Also assume that for the same instant $t$ the points x´´s of R´ (ST(S, S2 U $S_d$), $\tau$, $\sigma$) for which $\Psi´r (x´, t) \neq 0$ and/or $\Psi´i (x´, t)) \neq 0$, where these are the real and imaginary components of the wave function of the particle $p$ in the reference frame R´, appear concentrated in a very small region in comparison to the region of R where the points x´s appear disperse. In this case we will say that the system of particles $S_d$ is a detector of particles.

In quantum mechanics the detection of an elementary particle, a photon for instance, is described as a process by which a delocalized photon in space is localized in the point of detection. Except for the photons having a very high frequency which corresponds to energies of the order the nuclear processes, the already mentioned point of detection is generally an atom or a molecule, which are extremely small in comparison with the size of macroscopic bodies. That is why in the scale of macroscopic descriptions of physics atoms may be viewed as points, but we must take into account two considerations. First, one may consider that an atom in any of its quantum states has the extension of the wave function describing this state. This would be in principle an infinite extension except when we restrict the atom or the molecule to be in a finite space with reflecting borders. Without this restriction the wave function will tend to zero as we get farther from the center of mass of the atom or the molecule, and will be strictly zero at the infinite. Then, an atom or a molecule can be considered to have a finite size only by imposing an additional criterion of threshold for the wave function amplitude, below which it is assumed that the wave function has zero amplitude. It is only in this sense that one may consider atoms and molecules having diameters going from 1 to 100 Å, and so that they are so small in comparison with a macroscopic body that they may be considered as points. The second remark is related to distances much smaller than the diameter of an atom such as arises in Quantum Theory of Fields, or for elementary particles according to the Standard Model, or still in String Theory, and more recently in M theory. In physics it is the distance of Planck that is considered as the more fundamental distance by its magnitude, which is $10^{-33}$ cm. The size of the diameter of an atom or a molecule is then 25 orders of magnitude larger than the Planck distance. Thus, when we say that a photon is localized by a detector, what we are saying is that a photon has been absorbed by an atom or a molecule of the detector, and that in this way it has been integrated to a physical entity having a diameter of an enormous size when compared to the Planck distance, or compared to the size of a string, or even to the diameter of a proton. Then we may say that when a photon is localized in a detection process, this corresponds in our model to that the photon wave function appears now in a region of a reference frame of a much more restricted size than the region of macroscopic size in which the photon wave function extended before the detection, but this small size restricted region may still corresponds to an enormous number of points of the reference frame where the detection is carried out. The same argument applies to the case of photons of a so high frequency corresponding to energies involved in nuclear processes. In this case the localization of the photon will correspond to a region much smaller than for the case of atoms and molecules, but still much larger than that of the Planck scale.

In order to get a more intuitive view of the process of detection of a microscopic particle, let us consider the case of a photon and a detector such that one of its components is a small plate covered of sodium atoms. We know that if the frequency associated to the photon coincides with a resonance for absorption of energy, it may occur that an atom make an allowed transition from one to another of its states, and in this way the photon is localized in this atom during a lapse of time of the order of the mean life time of the state to which the atom has been excited. Depending on the detector, the mentioned localization may produce a cascade of electrons which is the macroscopic signal of the photon detection. When the photon is localized in one atom, the amplitude of wave function of the photon becomes almost zero except in the region where the wave function associated to the state to which the atom has been excited has sizable values. If now we make use of our definition of the detection of a microscopic particle, what we have just described may be seen in the following way: the photon has in the reference frame R and at the instant t zones where $N+ > N-$ and other zones where $N+ < N-$. These zones will correspond respectively to positive and negative amplitudes of the real component of the wave function in the instant t. The same will occur with the imaginary component for the cases $T+ > T-$ and $T+ < T-$. A photon delocalized in a reference frame R may be the result that a particle *p* may enter in a very large number of points of the reference frame R, and that many of these points may be distributed dispersedly in R so that they are separated by large distances. A particle *p* having many alpha-states may intercept different points of crossing when non empty intersection of different alpha-state of this particle occurs with different centers of points of crossing. These points of crossing may have different structures and therefore they belong to different points of R. Moreover, the same alpha-state of a particle *p* may have non empty intersections with the centers of different points of crossing entering in different points of R. On the other hand, points of crossing in which enter the particle *p* may belong to points which are distant from each other in the reference frame R, since different points of R may have very different space-time coordinates in R. Depending on the number of preparticles belonging to the alpha-states of the particle *p,* the points of R in which enter the particle p may be few, either localized or delocalized, when this number of preparticles is relatively small. In the complementary case, the number of points of R in which enter the particle *p* could be very large, and again localized or delocalized. A detector $S_d$ of particles has the property of tending equalize N+ with N- and T+ with T- in a temporal range Δt larger than a threshold for all the points of R except for a restricted region of R. This may occur by a redistribution of points of crossing among the points of R when considering the system of observation S2´=S2 U $S_d$ instead of S2. For points of R such that *N+(x, t)* — *N-(x, t)* and *T+(x, t)* — *T-(x, t)* are near to zero, it would be more probable that these values remain near to zero when the redistribution of points of crossing occurs in passing from the observation system S2 to the observation system S2 U $S_d$. On the other side, when the above mentioned differences are very different from zero then these differences will tend to remain near to these high values.

SPACE-TIME AND QUANTUM STATISTIC OF PARTICLES

Starting from the concepts of space-time and reference frame, and of a definition of the state of a particle, we now analyze the problem of quantum statistics in the framework of our theory. This problem was analyzed in reference 31, and here we will examine some aspects considered in this reference.

Consider a particle is represented by a set given by equation 2. We will designate particles represented by such sets, whose elements are in turn sets of preparticles, as particles of class 1. On the other hand, we will consider another sort of particles that we will designate as particles of class 2, which are related to the topology of space-time, i.e. related to the way in which the points of a space-time are connected. These particle of class 2 correspond to cuts or rips in a space-time. We will show that these two classes of particles are such that the class 1 obey the Bose-Einstein statistic, and that of class 2 the statistic of Fermi-Dirac (31). In order to see it we need a precise definition of trajectory and state of a particle in a given reference frame.

In pages 23-24 we have defined the trajectory $T(R(ST(S, S2), \tau, \sigma), p_j)$ of a particle $p_j$ of class 1 in the reference frame $R(ST(S, S2), \tau, \sigma)$ as a set of space-time coordinates ordered according to the increasing value of the temporal coordinate of the points of this reference frame for which alpha-states of $p_j$ yield a non empty intersection with centers of points of crossing belonging to these points of $R(ST(S, S2), \tau, \sigma)$. In analogy, we define the trajectory $T(R(ST(S, S2), \tau, \sigma), p_j)$ of a particle $p_j$ of class 2 in the reference frame $R(ST(S, S2), \tau, \sigma)$, as the ordered set of space-time coordinates of the points of this reference frame that belong to the borders of the cuts or rips corresponding to the particle $p_j$. The total or partial order of the elements of the trajectory $T(R (ST(S, S2), \tau, \sigma), p_j)$ is determined according to the increasing value of the temporal coordinate.

Definition 4: Consider the space-time $ST(S, S2)$ with an extension defined with respect to a given system of observation S2, and a reference frame $R (ST(S, S2), \tau, \sigma)$ defined in this space-time. The state of a particle $p$ of class 1 in this space-time $ST(S, S2)$ is given by the track of this particle in this space-time, and we denote this state $\varepsilon 1ST(ST(S, S2), p)$. In analogy, the state of a particle p of class 1 in the reference frame $R (ST(S, S2), \tau, \sigma)$ is specified by the trajectory of the particle $p$ in this reference frame, and we denote it $\varepsilon 1R (T(R (ST(S, S2), \tau, \sigma), p)$.

Definition 5: given a particle $p´$ of class 2 in the space-time $ST(S, S2)$ having an extension defined with respect to the system of observation S2, the state of $p´$ in this space-time is given by the points of $ST(S, S2)$ corresponding to the borders of the cuts or rips that represent the particle $p´$ in $ST(S, S2)$. We denote this state as $\varepsilon 2ST (ST(S, S2), p´)$. In analogy to the case of particles of class 1, the state of a particle of class 2 in the reference frame $R(ST(S, S2), \tau, \sigma)$, is given by the trajectory of the particle p´ in this reference frame, and we denote it $\varepsilon 2R(T(R(ST(S, S2), \tau, \sigma), p´)$.

These definitions are consisting with the postulate 2 according to which the points of a space-time distinguish from each other only by the structure of each of them. The greatest specification that can be given of a particle of either class 1 or class 2 in a space-time is thus the specification of the points belonging to the tracks of these particles in this space-time.

Concerning the particles of class 1, we have: (i) Given a particle of class 1 in the space-time T(S, S2), to this particle will correspond only one state in this space-time; (ii) Given the state $\varepsilon_1$(ST(S, S2), p) of a particle p in the space-time ST(S, S2), this state determines completely the trajectory of this particle in any reference frame R(ST(S, S2), $\tau$, $\sigma$), i.e. to a particle of class 1 corresponds only one trajectory in each reference frame R(ST(S, S2), $\tau$, $\sigma$), and the trajectories of the same particle in two different referent frames may be different between them; (iii) Many particles of class 1 may correspond to the same state in a given space time.

To illustrate the above point (iii), let us consider two particles *p* and *p´* of class 1 fulfilling that the particle p enters in a point of crossing that belongs to a point of the space-time ST(S, S2), and particle p´ enters in at least a point of crossing belonging to the same point of the space-time ST(S, S2), and the same occurs when we consider the reverse situation from p´ to p. When a very large number of particles enter in the system S, from the definition of particle of class 1 we see that there is no restriction for the mentioned situation to occur. A particle of class 1 enters in a point of crossing that belongs to a point of a space-time when at least one of its alpha-state has a non empty intersection with this point of crossing, and this may occur for many particles of class 1, since these particles may enter in the same point of crossing or in different points of crossing belonging to the same point of the space-time considered. Therefore, according to the point (iii), for a reference frame defined in a space-time ST(S, S2) where the system S has a very large number of particles, many particles of class 1 may have the same trajectory in this reference frame.

According to (i)-(iii), it can be shown that particles of class 1 fulfill the Bose-Einstein statistic. To see it let us consider that the reference frame where the particles of class 1 are described is an inertial reference frame (see the section: "Non inertial reference frames and gravitation", in page 24). In Section 3 of reference 29 we have shown that it can be ascribed values to the energy of the trajectories corresponding to well defined values of velocity. Then, consider the system S= { $p_1$, $p_2$,...,$p_N$} of N particles of class 1 described in an inertial reference frame R. Taking into account the properties (i) - (iii) of particles of class 1 already mentioned, we are immediately led to the partition function of the system S given by (31):

$$Z = \sum_{n_1, n_2, \ldots, n_s} \exp\left[-\beta(n_1\varepsilon_1 + n_2\varepsilon_2 + \ldots + n_s\varepsilon_s)\right], \qquad (7)$$

Where $n_j$ is the number of particles of S in the state j, $\varepsilon_j$ is the energy corresponding to the state j in the reference frame R (ST(S, S2), $\tau$, $\sigma$), and $\sum n_j = N$, where N is the total number of particles of class 1 described in the reference frame R. In equation 7 we have taken into account that each state j corresponds to only one trajectory in the inertial reference frame R(ST(S, S2), $\tau$, $\sigma$), and consequently also corresponds to only one value of energy $\varepsilon_j$. On the other hand, in equation (7) does not appear a factor $N! / n_1! \, n_2! \ldots n_s!$ in each term of the summation. The inclusion of this factor would take into account all possible ways in which the particles $p_i$, $i = 1, N$, may be found in the one particle states $j=1, S$, in which case equation 7 leads to the Maxwell-Boltzmann classical distribution. On the other hand, if we do not include the factors $N!/n_1!\,n_2!\ldots n_s!$ in equation 7, as must be done in our case of particles of class 1, then equation (7), with the restriction $\sum n_j = N$, leads immediately to the Bose-Einstein distribution (42).

Consider now the case of particles of class 2. In this case we represent the state $\varepsilon 2ST(ST(S, S2), p´)$ by a set of subsets $C_k$ of points of $ST(S, S2)$ such that the elements of $C_k$ are the points of this space-time that define the borders of the k- cut corresponding to the particle p´. When to the particle p´ corresponds M cuts in the space-time $ST(S, S2)$, we may specify its state by the set:

$$\varepsilon 2ST(ST(S, S2), p´) = \{C_1, C_2,...,C_M\} \qquad (8)$$

Now, particles of class 2 fulfill a similar property to the property (i) above mentioned, i. e. we have the property: (j) given a particle of class 2 and a space-time $ST(S, S2)$, to this particle corresponds only one state $\varepsilon 2ST(ST(S, S2), p´)$ in this space-time. We have also a similar property to that mentioned in (ii), which is: (jj) given a state $\varepsilon 2ST(ST(S, S2), p)$ of a particle p´ of class 2 in a space-time $ST(S, S2)$, this state determine completely the trajectory of this particle in any reference frame $R(ST(S, S2), \tau, \sigma)$. I.e. to each particle of class 2 corresponds only one trajectory in each reference frame $R(ST(S, S2), \tau, \sigma)$, and the trajectories in two different reference frames may be different to each other. Yet, particles of class 2 do not fulfill a similar property to the one above mentioned in (iii), concerning particles of class 1, since now we have: (jjj) no more than one particle of class 2 may correspond to the same state in a given space-time $ST(S, S2)$, and to the same trajectory in any reference frame $R(ST(S, S2), \tau, \sigma)$.

According to (j) and (jj), and that the state $\varepsilon 2ST(ST(S, S2), p´)$ of a particle of class 2 in the space-time $ST(S, S2)$ determines completely the trajectory of p´ in any reference frame $R(ST(S, S2), \tau, \sigma)$, we have that to each particle of class 2 corresponds only one trajectory in each reference frame $R(ST(S, S2), \tau, \sigma)$. However, there is an important difference with the case of particles of class 1 which is related to point (jjj) specified above, according to which two particles p´ and p´´ of class 2 in a space-time $ST(S, S2)$ cannot be found in the same state in this space-time, and consequently cannot have the same trajectory in any reference frame $R(ST(S, S2), \tau, \sigma)$. In order to see it, let us assume that p´ and p´´ are in the same state $\{C_1, C_2,...,C_M\}$ (eq. 8). Since every $C_k$ specifies completely the points defining the k-cut in $ST(S, S2)$, each point belonging to the border of this cut is different by its structure from any other point of $ST(S, S2)$. Therefore the two particles p´ and p´´ will correspond to the same cuts in $ST(S, S2)$, and so we have p´= p´´. I.e. in a state *j* as described in equation 8 it cannot be found more than one particle, and consequently there is no more than one particle of class two with the same trajectory in a reference frame $R(ST(S, S2), \tau, \sigma)$. Then, taking into account the properties (j) - (jjj) for particles of class 2, these particles must fulfill equation 7 with the restrictions $n_1 = n_2 = \ldots = n_s = 1$ and $\sum n_j = N$, which leads immediately to the Fermi-Dirac statistic (42).

This result indicates that the difference among Maxwell-Boltzmann, Bose-Einstein and Fermi-Dirac statistics, is related to the assumption implicit in the Maxwell-Boltzmann distribution according to which what we consider to be a particle corresponds in fact to a collection of particles (31). To clarify this point, let us consider a system of macroscopic particles $p_1, p_2,..., p_r$. According to the classical description, each one of these particles may be found in different states and a particle change of state when it interacts with other particles, or with a field, and this interaction is mediated by the exchange of other particles. For instance, consider the case of the interaction between particles having electric charge. This interaction is mediated by the exchange of photons between charged particles. Thus, when we consider that a macroscopic particle $p_r$ may occupy different states, in fact we are considering $p_r$ as a system of particles the number of which may change. In the classical description one indentifies the system $p_r$ as it would consist of only one macroscopic particle that may change of state. If one identifies macroscopic particles in this way and after that one makes the corresponding counting of them, and of the states in which these particles may be found, then one must take into account a factor $N!/ n_1! \, n_2! \ldots n_s!$ in each term of the summation appearing in equation 7. This leads to the Maxwell-Boltzmann distribution which describes correctly the statistic for systems of macroscopic particles such that each of these particles is interpreted in the classical way we have just considered.

In the case of the quantum description of system of microscopic particles, as it is usually done, one again interprets each particle as a unity that may change of state, and similarly to what we have explained in the case of classical macroscopic particles, one counts the number of particles and the states in which these particles may be found. But then, one must take into account the indistinguishable character of these particles in order to get the statistics of Bose-Einstein for Bosons and of Fermi-Dirac for fermions. It is just by taking into account the indistinguishable character of the particles that one erases the effect of the factors $N! / n_1! \, n_2! \ldots n_s!$, that according to this point of view should multiply each term of the summation in equation 7 (31).

Finally, according to our theory, a microscopic particle cannot have different states in a given space-time, and consequently its trajectory in each reference frame corresponds to only one energy. As we have shown above, this leads to the Bose-Einstein statistics when different particles may be found in the same state (particles of class 1), and to the Fermi-Dirac statistics when no more than one particle may be found in a given state (particles of class 2). Therefore, our description of bosons and fermions requires simpler assumptions than the usual ones for obtaining a correct description of the quantum statistics.

CONCLUSIONS

We have analyzed a fundamental theory of time and space-time starting from the primitive concepts of preparticle and of the membership relation of set theory. We have considered four postulates consisting with the idea according to which what allows distinguishing points in a space-time among them is the structure of points of crossing of particles: equivalent classes of points of crossing with the same structure represent the points of a space-time. We have described as derivative concepts those of time, space-time, reference frame, particle, field, and interaction between fields. Two classes of particles are considered: particles of class 1 represented by sets of subsets of preparticles, and particles of class 2 represented by cuts or rips in a space-time. It is shown that particles of class 1 fulfills the Bose-Einstein statistics, and those of class 2 the statistics of Fermi-Dirac. We also describe as derivative concepts those of wave-function and detector of particles. We thus see that starting from relatively simple primitive concepts, and from postulates that give rise to a simple model of the structure of space-time, one is able to describe several fundamental properties of the physical word. On the other hand, it is clear that the theory presented here has not too much "physical meat", and therefore much of it must be added to our theory to get a description of the physical reality. However, it is suggestive that a simple theory of space-time as that we have described here, allows the description of some of the fundamental properties of the physical world.

REFERENCES

1)Poincaré, H. "Les geometries non Euclidienes", Revue générale des Sciences pures et appliquées, **2,** 669-774 (1891).

2)Wheeler, J. A. Geometrodynamics, Academia Press, New York. 1962.

3)Wheeler, J. A., "Quantum Theory and Gravitation" (ed. A.R. Marlow) Academic Press, New York. 1980.

4)Patton, C. M. , Wheeler, J. A. Is Physics Legislated by Cosmogony. "Quantum Gravity", an Oxford Symposium, Edited by Isham, C.J., Penrose, R., and Sciama, D.W., Clarendon Press. Oxford, 1975, 538-605.

5)Bunge, M., "Philosophy in Crisis: The Need for Reconstruction". Published by Prometheus Books, New York, page 18 (2001).

6)Penrose, R., "Angular Momentum: an approach to combinatorial space-time", en Quantum Theory and Beyond (T. Bastin, ed.), Cambridge U. Press, 1971.

7)Penrose, R., "On the Nature of Quantum Geometry", en Magic without magic (J. R. Klauder ed.) Freeman, New York, 1972.

8)Sparling G., Homology and Twistor Theory, "Quantum Gravity", an Oxford Symposium, Edited by Isham, C.J., Penrose, R., and Sciama, D.W., Clarendon Press. Oxford, 1975, 408-499.

9)Dewitt, B. S., Phys. Rev. 162, 5, 1195 (1967).

10)Feynman, R., Acta Physica Polonica, XXIV, 697 (1963).

11)Isham, C. J., An Introduction to Quantum Gravity, "Quantum Gravity", an Oxford Symposium, Edited by Isham, C.J., Penrose, R., and Sciama, D.W., Clarendon Press. Oxford, 1975, 1-77.

12) Duff, M. J., Covariant Quantization, "Quantum Gravity", an Oxford Symposium, Edited by Isham, C.J., Penrose, R., and Sciama, D.W., Clarendon Press. Oxford, 1975, 78-135.

13) Deser, S., Quantum Gravitation: Trees, Loops and Renormalization, "Quantum Gravity", an Oxford Symposium, Edited by Isham, C.J., Penrose, R., and Sciama, D.W., Clarendon Press. Oxford, 1975, 136-170.

14) Salam, A., Impact of Quantum Gravity Theory on Particle Physics, "Quantum Gravity", an Oxford Sym0posium, Edited by Isham, C.J., Penrose, R., and Sciama, D.W., Clarendon Press. Oxford, 1975, 500-537.

15) Greene B., "The Elegant Universe", Vintage Books. A Division of Random House, Inc., New York, 1999. Caps. 3 and 7.

16) Bunge M., "Controversias en Física", Editorial Tecnos, S. A., Madrid, 1983.

17) Bunge, M., "Foundations of Physics", Springer-Verlag, New York, 1967.

18) Eagleman, D. M. "El Tiempo y el Cerebro" en "La Ciencia del Futuro", Max Brockman, Ed. , RBA, Barcelona, 2010.

19) Noll, W., Space-Time Structure in Classical Mechanics, Delaware Seminar in the Foundations of Physics, pp. 28-34, Springer-Verlag, New York, 1967.

20) Bunge, M., "Philosophy of Science", 35, 355 (1968).

21) Bunge, M., "Studium Generale", 23, 562 (!970).

22) Lorente, M. "Quantum Processes and the Foundation of Relational Theories of Space and Time". In Proceedings of the Relativity Meeting`93: Relativity in General (Eds. Diaz Alonso, J. and Lorente Paramo, M). Editions Frontieres, Gif-sur-Ivette, France, 1994. pp.297-302.

23) Lorente, M., "A Realistic Interpretation of Lattice Gauge Theory". En "Fundamental Problems in Quantum Physics (Eds. Ferrero, M. and van der Merwe, A.) Kluwer Academic Publishers, Dordrecht, 1995. pp. 177-186.

24) Hilbert, D. Grundlage der Geometrie (Teubner, Leipzig, 1899). Traducción al español: "Fundamentos a la Geometría" (C.S.I.C. , Madrid, 1991). Traducción al ingles: "The Foundations of Geometry". Merchant Books, USA.

25) Lorente, M. "Modernas Teorías sobre el tiempo discreto". En: "El Tiempo: tiempo relatividad y saberes", Publicaciones de la Universidad Pontificia Comillas, Madrid, 1995. pp. 199-206.

26) García-Sucre, M., International Journal of Theoretical Physics 12, 25 (1975).

27) García-Sucre, M., International Journal of Theoretical Physics 17, 163 (1978).

28) García-Sucre, M., "Proceedings of the First Section of the Interdisciplinary Seminars on Tachyons, Monopoles and Related Topics", E. Recami (ed.), North Holland, Amsterdam, 1978. pp. 235-246.

29) García-Sucre, M., International Journal of Theoretical Physics 18, 725 (1979).

30) García-Sucre, M., "A Kind of Collapse in a Simple Spacetime Model", in Scientific Philosophy Today, J. Agassi and R.S. Cohen (eds), D. Reidel Publishing Company, 1981, pp. 45-69.

31) García-Sucre, M. International Journal of Theoretical Physics 24, 441 (1985).

32) Greene B., "The Fabric of the Cosmos: space, time and the texture of reality", published by Alfred A. Knopf, New York, 2011, Cap. 15, pp 452-455.

33) Russell B., "Wisdom of the West", Ed. Paul Foulkes, Crescent Books. Inc., Rathbone Books, London, 1959.

34) Jammer M., "Concepts of Space: the History of Theories of Space in Physics", Dover Publications, INC, Mineola, New York, 1954.

35) Greene B., "The Fabric of the Cosmos: space, time and the texture of reality", published by Alfred A. Knopf, New York, 2011, Caps. 1 and 4.

36) Leibniz, G., W. "Tres Textos Metafísicos". Editorial Norma S. A., Santa Fe de Bogotá, Colombia, 1992.

37) Leibniz, G., W. "Discourse on Metaphysics and other Essays" (Edited and Translated by Garber, D. and Ariew, R.). Hackett Publishing Company, Indianapolis and Cambridge, 1991.

38) Einstein A., Podolsky Y, Rosen N., Physical Review, 47,777 (1935).

39) Aspect, A., Grangier, P, Roger, G., Physical review Letters, 49, 91 ,1982.

40) Scully, M. O., Drühl, K., Physical Review, A 25, 2208 (1982).

41) Zeh, H. D., "Basic concepts and their interpretation", en "Decoherence and the appeareance of a classical world in quantum theory", Eds, Joos, E., Zeh, H. D., Kiefer , C., Giulini, D., Kupsch, J., Stamatescu, I.O., Springer-Verlag, Berlin, Heidelberg, New York,

1996, 2003. pp 7-40.

42)Reif, F., "Fundamentals of statistical and thermal physics, Chapter 9, McGraw-Hill, New York, 1965.

$\subset$

$>$